\def\be{\begin{equation}}
\def\ee{\end{equation}}
\def\bea{\begin{eqnarray}}
\def\eea{\end{eqnarray}}
\def\nn{\nonumber}
\def\qtwo{\qquad\qquad}
\def\qthree{\qquad\qquad\qquad}
\def\qfour{\qquad\qquad\qquad\qquad}

\def\threej#1#2#3#4#5#6{\left( \begin{array}{ccc} #1 & #2 & #3 \\ #4 & #5 & #6 \end{array} \right) }
\def\eqdef{\stackrel{\rm def}{=}}
\def\bigoh{{\mathcal O}}
\def\Cov{\mbox{Cov}}
\def\Var{\mbox{Var}}
\def\Tr{\mbox{Tr}}

\def\estE{{\mathcal E}}
\def\bk{{\bf k}}
\def\bx{{\bf x}}
\def\symm{{+\mbox{ (symm.)}}}
\def\perm{{+\mbox{ (perm.)}}}

\def\Nband{N_{\rm band}}
\def\Npix{N_{\rm pix}}
\def\Nfact{N_{\rm fact}}
\def\Nopt{N_{\rm opt}}
\def\Nquad{N_{\rm quad}}
\def\Ntheta{N_\theta}
\def\Nphi{N_\varphi}

\def\ellmin{\ell_{\rm min}}
\def\ellmax{\ell_{\rm max}}

\def\fsky{f_{\rm sky}}

\def\simle{\lesssim}

\def\bps{b^{\rm ps}}
\def\fnlloc{f_{NL}^{\rm loc}}
\def\Floc{F^{\rm loc}}
\def\bloc{b^{\rm loc}}
\def\Bloc{B^{\rm loc}}
\def\fnlgrav{f_{NL}^{\rm grav}}
\def\Fgrav{F^{\rm grav}}
\def\bgrav{b^{\rm grav}}
\def\fnlhd{f_{NL}^{\rm hd}}
\def\Fhd{F^{\rm hd}}
\def\bhd{b^{\rm hd}}
\def\fnleq{f_{NL}^{\rm eq}}
\def\Feq{F^{\rm eq}}

\documentclass[twocolumn,amsmath,amssymb,prd]{revtex4}
\usepackage{bm}
\usepackage{epsf}
\usepackage{wick}

\begin{document}

\title{Algorithms for bispectra: forecasting, optimal analysis, and simulation}
\author{Kendrick M. Smith}
\affiliation{Kavli Institute for Cosmological Physics, University of Chicago, 60637}
\affiliation{Department of Physics, University of Chicago, 60637}
\author{Matias Zaldarriaga}
\affiliation{Institute for Theory and Computation, Harvard-Smithsonian Center for Astrophysics,
MS-51, 60 Garden Street, Cambridge, MA 02138, USA}
\affiliation{Jefferson Laboratory of Physics, Harvard University, Cambridge, Massachusetts 02138, USA}

\begin{abstract}
\baselineskip 11pt
We propose a factorizability ansatz for angular bispectra which permits fast algorithms for
forecasting, analysis, and simulation, yet is general enough to encompass many interesting CMB
bispectra.  We describe a suite of general algorithms which apply 
to any bispectrum which can be represented in factorizable form.  First, we present algorithms
for Fisher matrix forecasts and the related problem of optimizing the factorizable representation, 
giving a Fisher forecast for Planck as an example.  We show that the CMB can give independent constraints 
on the amplitude of primordial bispectra of both local and equilateral shape as well as those created by 
secondary anisotropies.  We also show that the ISW-lensing bispectrum should be detected by Planck and could bias 
estimates of the local type of non-Gaussianity if not properly accounted for.  
Second, we implement a bispectrum estimator which 
is fully optimal in the presence of sky cuts and inhomogeneous noise, extends the generality
of fast estimators which have been limited to a few specific forms of the bispectrum, and improves 
the running time of existing implementations by several orders of magnitude.
Third, we give an algorithm for simulating random, weakly non-Gaussian maps with prescribed power 
spectrum and factorizable bispectrum.
\end{abstract}

\maketitle

\section{Introduction}

Despite the rapid progress in observational cosmology in the last decade,  there are very few observational probes 
that are able to constrain the first instants in the evolution of the Universe, when density perturbations were created. 
In addition to the shape of the primordial spectrum of perturbations and the amplitude of the gravitational wave 
background, any measurable departure from pure Gaussianity in the statistics of the primordial seeds would severely 
constrain the inflationary dynamics. The level of non-Gaussianity in the simplest, single field, inflationary models 
has now been robustly calculated and shown to be too small to be measured by upcoming Cosmic Microwave Background (CMB) 
experiments \citep{Maldacena:2002vr,Acquaviva:2002ud}. Various departures from the simplest scenario however are thought 
to produce observable signals \citep{Zaldarriaga:2003my,Creminelli:2003iq,Arkani-Hamed:2003uz,Alishahiha:2004eh,Rigopoulos:2005ae,Rigopoulos:2005us,Chen:2006nt,Chen:2006xj}. 

In fact it has become clear that the three point function of the primordial fluctuations, the bispectrum, is the most 
promising statistic to probe the small departures from Gaussianity that could originate during 
inflation \citep{Creminelli:2006gc}. Moreover the structure of the three point function contains precious information 
about the inflationary dynamics \citep{Babich:2004gb}. For example as a result of causality, the three point function 
in models where only one field is effectively responsible for inflation have to satisfy strict consistency relations that 
constrain the configurational dependance of the bispectrum \citep{Maldacena:2002vr, Creminelli:2004yq}.  
Furthermore we have learned that the shape of the bispectrum of any primordial non-Gaussianity arising in inflationary 
models neatly falls in two separate classes.  The  momentum space three point function in single field models is largest 
when the three momenta are of similar magnitude, while for multi-field models  where curvature perturbations are 
created by a field different from the one that drives inflation, the bispectrum is largest in the squeezed limit, 
when one of the momenta is much smaller than the other two. 

If we are to probe the primordial non-Gaussianities using the CMB we also need to consider the departures from 
Gaussianity produced by secondary anisotropies as well as residual foreground contamination 
({\it eg.} \citep{Spergel:1999xn,Goldberg:1999xm,Verde:2002mu,Cooray:1999kg}). In general the non-linear dynamics of 
Gravity and of any probe we wish to use will lead to some departures from Gaussianity which are not of primordial origin.
These additional sources on non-Gaussianity produce bispectra with specific configurational dependance. 

It is clear then that if we want to hunt for possible departures from Gaussainity as well as constrain the evolution of 
perturbations after decoupling through secondary anisotropies, we need to develop data analysis tools that will allow us 
to measure the bispectrum efficiently and can distinguish between different shapes of the momentum space bispectrum. 
The efficiency of the tools will be crucial as the expected level of non-Gaussianity is rather small so it will only be 
detectable in large surveys, with a large number of pixels. Developing these tools is the object of this paper. 
The effort is timely as many of the predicted signals are expected to be detectable by the upcoming Planck satellite. 

In spite of the promise offered by the three-point function, and the variety of forms that have
been calculated, there is currently a lack of general methods for connecting a predicted 
three-point function with data.
The basic problem is that the most general three-point function allowed by rotational invariance
is described by its angular bispectrum $B_{\ell_1\ell_2\ell_3}$, an object with $\bigoh(\ellmax^3)$
degrees of freedom.
In this generality, algorithms tend to be prohibitively expensive; for example, evaluating an optimal
bispectrum estimator has cost $\bigoh(\ellmax^5)$.
Computationally feasible algorithms are only known for a few special forms of the bispectrum \citep{Komatsu:2003iq,Creminelli:2005hu}.
One might make an analogy with power spectrum estimation; many solutions have been proposed to the problem
of optimally estimating power spectra from data
\citep{Bond:1998zw,Wandelt:2001fz,Oh:1998sr,Wandelt:2000av,Szapudi:2000xj,Hivon:2001jp,Efstathiou:2003dj,Wandelt:2003uk,Jewell:2002dz},
but given a prediction for the shape of the bispectrum which is to be estimated from data,
relatively little is known.

The purpose of this paper is to address this problem by proposing a ``toolkit'' of
algorithms for forecasting, optimal estimation, and non-Gaussian simulation.
These fast algorithms will apply to any bispectrum which can be represented in a factorizable form which
will be defined below (Eq.~(\ref{eq:factdef})).
We will show (\S\ref{sec:fact}) that a wide range of previously calculated CMB bispectra can be represented in this form,
thus giving wide scope to our methods.
Our methods do not apply to a completely arbitrary bispectrum $B_{\ell_1\ell_2\ell_3}$, but we believe this to
be a necessary feature of any treatment which leads to fast algorithms.
The factorizability criterion is a compromise in generality which is specific enough to enable practical computation,
yet general enough to encompass a large number of interesting cases.

Our first algorithm (\S\ref{sec:fisher}-\ref{sec:optnfact}) attempts to ``optimize'' a bispectrum by reducing the size of its
factorizable representation.  We will see that this is closely related to computing a Fisher matrix forecast
under the assumption of homogeneous noise.  The optimization algorithm can be used as a preliminary step to
speed up the other algorithms; we will also see examples where a bispectrum with an intractably large
factorizable representation is optimized down to manageable size, thus giving additional scope
to our methods.  As an application, we will present (\S\ref{sec:fisherforecasts}) a multi-bispectrum 
Fisher matrix forecast for Planck.

Second, we give a general Monte Carlo framework for estimating bispectra from noisy data, in the
presence of sky cuts and inhomogeneous noise (\S\ref{sec:est}-\S\ref{sec:T}).
This generalizes the estimator proposed by Komatsu, Spergel, and Wandelt \citep{Komatsu:2003iq} to arbitrary factorizable
bispectra, and also improves it in the case of inhomogenous noise, by including the linear term in the estimator
proposed in \citep{Creminelli:2005hu}.
We also present some code optimizations which dramatically improve existing implementations and make Monte
Carlo estimation practical for Planck.

Third, we give a simulation algorithm (\S\ref{sec:ngsim}) which outputs random non-Gaussian maps with arbitrary power spectrum
and factorizable bispectrum.  This greatly extends the generality of existing methods for simulating non-Gaussian
fields \citep{Komatsu:2003fd,Liguori:2003mb,Contaldi:2001wr,Rocha:2004ke}.

Throughout this paper, we make the assumption of weak non-Gaussianity.
On a technical level, this means that the covariance of the three-point function is well
approximated by its Gaussian contribution.  If this approximation breaks down, then both
the choice of optimal estimator and the estimator variance can be affected.  It might
seem that the CMB is so close to Gaussian that this is never an issue; however, it has
recently been shown \citep{Creminelli:2006gc} that it can be important for bispectra of ``squeezed'' shape,
if a several-sigma detection can be made, even though the field is still very close to Gaussian. 
A second reason that the assumption of weak non-Gaussianity might break down is that the
lensed CMB is non-Gaussian; e.g. in \cite{Babich:2004yc}, it is argued that this may degrade constraints
on $f_{NL}$ by $\sim 25\%$ at Planck sensitivity.
Both of these effects are outside the scope of this paper.

\section{Notation and conventions}
\label{sec:notation}

\par\noindent
The angular CMB bispectrum $B_{\ell_1\ell_2\ell_3}$ is defined by
\be
\langle a_{\ell_1 m_1} a_{\ell_2 m_2} a_{\ell_3 m_3} \rangle = 
B_{\ell_1\ell_2\ell_3} \threej{\ell_1}{\ell_2}{\ell_3}{m_1}{m_2}{m_3}.
\label{eq:Bdef}
\ee
where the quantity in parentheses is the Wigner 3j-symbol.
This is the most general three-way expectation value which is invariant under rotations.
Following \citet{Komatsu:2001rj}, we define the reduced bispectrum $b_{\ell_1\ell_2\ell_3}$ by
\bea
B_{\ell_1\ell_2\ell_3} &=& \left[ \frac{(2\ell_1+1)(2\ell_2+1)(2\ell_3+1)}{4\pi} \right]^{1/2}               \nn \\
                        && \qtwo \times  \threej{\ell_1}{\ell_2}{\ell_3}{0}{0}{0}  b_{\ell_1\ell_2\ell_3}.    \label{eq:bdef}
\eea
Because the 3j-symbol on the right-hand side vanishes for triples $(\ell_1,\ell_2,\ell_3)$
such that $(\ell_1+\ell_2+\ell_3)$ is odd, Eq.~(\ref{eq:bdef})
only makes sense if $B_{\ell_1\ell_2\ell_3}$ also vanishes for such triples.
This condition is equivalent to parity
invariance of the bispectrum, and will be satisfied for all bispectra considered in this paper.

\section{Factorizability}
\label{sec:fact}

\par\noindent
A basic problem in the theory of bispectrum estimation is that there are so many degrees of
freedom in the bispectrum $B_{\ell_1\ell_2\ell_3}$ that completely general methods, which make no assumption
on the form of $B_{\ell_1\ell_2\ell_3}$, are computationally prohibitive.
For example, even with the unrealistic assumption of homogeneous instrumental noise, the cost
of evaluating an optimal estimator is $\bigoh(\ellmax^5)$ \citep{Komatsu:2002db}.

In this section, we propose a factorizability ansatz for the form of the bispectrum, and show that many
CMB bispectra of theoretical interest can be written in factorizable form.
Our approach is empirical; we simply collect and analyze interesting bispectra from the literature.
In subsequent sections, we will present fast algorithms which can be applied to factorizable bipsectra,
which will improve the $\bigoh(\ellmax^5)$ cost and make calculations tractable.

Our ansatz is that the {\em reduced} bispectrum defined in Eq.~(\ref{eq:bdef}) is a sum of terms factorizable
in $\ell_1$, $\ell_2$, $\ell_3$:
\be
b_{\ell_1\ell_2\ell_3} = \frac{1}{6} \sum_{i=1}^{\Nfact} X^{(i)}_{\ell_1} Y^{(i)}_{\ell_2} Z^{(i)}_{\ell_3} \symm
\label{eq:factdef}
\ee
where $\Nfact$ is not too large.  
In Eq.~(\ref{eq:factdef}) and throughout the paper, $\symm$ stands for the sum of five additional terms obtained
by permuting $\ell_1$, $\ell_2$, $\ell_3$.

\subsection{CMB secondaries}

The first general class of bispectra which satisfy the factorizability condition (Eq.~(\ref{eq:factdef})) are those
which arise from three-way correlations between the primary CMB, lensed CMB, and secondary anisotropies
\citep{Spergel:1999xn,Goldberg:1999xm}.
These are of a manifestly factorizable (with $\Nfact=3$) form,
\bea
b_{\ell_1\ell_2\ell_3} &=& \frac{\ell_1(\ell_1+1) - \ell_2(\ell_2+1) + \ell_3(\ell_3+1)}{2} C_{\ell_1} \beta_{\ell_3}  \nn \\
&& \qfour \symm  \label{eq:secbispec}
\eea
where $\beta_\ell$ depends on the secondary anisotropy which is considered.

For example, with secondary anisotropy given by the integrated Sachs-Wolfe (ISW) effect,
$\beta_\ell$ is equal to $C_\ell^{T\phi}$, the cross power spectrum between the unlensed CMB temperature and lensing potential.
In this case, the bispectrum should be detectable by Planck \citep{Goldberg:1999xm,Hu:2000ee},
and would provide a direct signature, internal to the CMB, of an evolving gravitational potential \citep{Seljak:1998nu}.
We will refer to this as the ISW-lensing bispectrum and use it as a running example throughout the paper.
Other examples of the general form in Eq.~(\ref{eq:secbispec}) have also been studied; e.g. the Rees-Sciama-lensing
bispectrum \citep{Verde:2002mu}, and the SZ-lensing bispectrum \citep{Goldberg:1999xm,Cooray:1999kg}.

Another general class of CMB bispectra is given by three-way correlations between Ostriker-Vishniac anisotropies
and secondary anisotropies \citep{Cooray:1999kg,Hu:1999vq}.  These bispectra are of the form:
\be
b_{\ell_1\ell_2\ell_3} = \int dr\, f_{\ell_1}(r) g_{\ell_2}(r) \symm
\ee
To make this factorizable, we replace the $r$ integral by a finite quadrature:
\be
b_{\ell_1\ell_2\ell_3} \rightarrow \sum_{i=1}^{\Nfact} (\Delta r_i) f_{\ell_1}(r_i) g_{\ell_2}(r_i) \symm
\ee
The bispectrum is then of the factorizable form (Eq.~(\ref{eq:factdef})), where $\Nfact$ is the
number of quadrature points needed to approximate the integral.
This device, replacing an integral by a finite sum in order to satisfy the factorizability condition,
will be used frequently throughout the paper.

Finally, we mention a mathematically trivial but practically important example: 
the bispectrum from residual point sources (assumed Poisson distributed) is given by
\be
\bps_{\ell_1\ell_2\ell_3} = \mbox{constant}   \label{eq:bps}
\ee
The value of $\bps$ will depend on the flux limit at which point sources can be detected and masked
in the survey considered, and on the assumed flux and frequency distribution of the sources.
As a rough baseline, motivated by \cite{Komatsu:2002db,Toffolatti:1997dk}, 
we will take $\bps_{\ell_1\ell_2\ell_3} = 10^{-8}$ $\mu$K$^3$ at Planck sensitivity,
corresponding to a flux limit $\sim 0.2$ Jy at 217 GHz.

\subsection{Primordial non-Gaussianity}

Moving on, we next consider CMB bispectra which arise from primoridal non-Gaussianity, rather than secondary anisotropies.
The general relation between the angular CMB bispectrum and primordial three-point function can be described as follows.
We assume that the primordial three-point function is invariant under translations and rotations, so that three-way
correlations in the Newtonian potential $\Phi$ are given by
\be
\langle \Phi(\bk_1) \Phi(\bk_2) \Phi(\bk_3) \rangle = (2\pi)^3 \delta^3(\bk_1+\bk_2+\bk_3) F(k_1,k_2,k_3)  \label{eq:3Ddef}
\ee
where $F$ depends only on the lengths $k_i = |\bk_i|$, as implied by the notation.
In \citep{Wang:1999vf}, it is shown that the {\em reduced} angular bispectrum is given by
\bea
b_{\ell_1\ell_2\ell_3} &=& \int dr\, r^2 
                           \left( \prod_{i=1}^3 \frac{2 k_i^2\, dk_i}{\pi} j_{\ell_i}(k_ir) 
                           \Delta^T_{\ell_i}(k_i) \right)                                       \nn \\
                        && \qthree \times F(k_1,k_2,k_3)                                        \label{eq:WK}
\eea
where $\Delta^T_{\ell_i}(k)$ denotes the transfer function between the CMB multipoles and the Newtonian potential:
\be
a_{\ell m} = 4\pi(i)^\ell \int \frac{d^3\bk}{(2\pi)^3} \Delta^T_\ell(k) \Phi(\bk) Y_{\ell m}^*(\hat\bk)
\ee

Now let us consider some specific examples of primordial non-Gaussianity.
The simplest is the ``local model'', in which the primordial potential satisfies:
\be
\Phi(\bx) = \Phi_G(\bx) - \fnlloc \big( \Phi_G(\bx)^2 - \langle\Phi_G(\bx)^2\rangle \big)
\ee
where $\fnlloc$ is a constant and $\Phi_G$ is Gaussian.  In this model, the primordial bispectrum is
\be
\Floc(k_1,k_2,k_3) = \fnlloc \left( \frac{\Delta_\Phi^2}{k_1^{4-n_s} k_2^{4-n_s}} \right) \symm
\label{eq:Flocal}
\ee
where $\fnlloc$ is a constant and $P^{\Phi}(k) = \Delta_\Phi/k^{4-n_s}$ is the primordial power spectrum.
Substituting Eq.~(\ref{eq:Flocal}) into Eq.~(\ref{eq:WK}), the angular CMB bispectrum is
\be
\bloc_{\ell_1\ell_2\ell_3} = \fnlloc \int dr\, r^2 \beta_{\ell_1}(r) \beta_{\ell_2}(r) \alpha_{\ell_3}(r) \symm
\label{eq:blocal}
\ee
where, following \citep{Komatsu:2001rj}, we have introduced the functions
\bea
\alpha_\ell(r) &=& \frac{2}{\pi} \int_0^\infty dk\, k^2 j_\ell(kr) \Delta^T_\ell(k)  \label{eq:alphadef} \\
\beta_\ell(r)  &=& \frac{2}{\pi} \int_0^\infty dk\, k^2 j_\ell(kr) \Delta^T_\ell(k) \left( \frac{\Delta_\Phi}{k^{4-n_s}} \right)
\eea
After replacing the $r$ integral in Eq.~(\ref{eq:blocal}) by a finite quadrature, the local bispectrum
is of factorizable form (\ref{eq:factdef}), with $\Nfact$ equal to the number of quadrature points.

This illustrates a general point: suppose that the primordial bispectrum $F(k_1,k_2,k_3)$ is a sum of $M$ terms 
each factorizable in $k_1$, $k_2$, $k_3$.
Then the resulting CMB bispectrum $b_{\ell_1\ell_2\ell_3}$ will be a sum of $\Nfact = M \Nquad$ terms each factorizable 
in $\ell_1$, $\ell_2$, $\ell_3$, where $\Nquad$ is the number of quadrature points need to do the $r$ integral in Eq.~(\ref{eq:WK}).

Our next example is taken from \citep{Liguori:2005rj}, in which a primordial bispectrum of the form
\bea
\Fgrav(\bk_1,\bk_2,\bk_3) &=& \frac{\Delta_\Phi^2}{k_1^{4-n_s} k_2^{4-n_s}} \fnlgrav(\bk_1,\bk_2,\bk_3)   \label{eq:Fgrav} \\
\fnlgrav(\bk_1,\bk_2,\bk_3) &=& -\frac{1}{6} - \frac{\bk_1\cdot\bk_2}{k_3^2} + \frac{3(\bk_1\cdot\bk_3)(\bk_2\cdot\bk_3)}{k_3^4} \nn
\eea
is studied, arising from second-order gravitational evolution after inflation \footnote{This form of the bispectrum 
is problematic as it does not satisfy a simple consistency condition resulting from causality 
requirements  \citep{Creminelli:2004yq,Creminelli:2004pv}.  The bispectrum does not vanish when one 
takes the limit of one mode being very large compared to the horizon. We will ignore this problem here 
as we are just taking it as an illustrative example that has been recently analyzed in detain in the 
literature. We will later find that this shape is very correlated with a linear combination of the local 
and equilateral shapes. The correlation with the local shape is probably unphysical, originating from the 
fact that this shape does not satisfy the appropriate consistency relation.}. 
Using the constraint $(\bk_1+\bk_2+\bk_3)=0$ from Eq.~(\ref{eq:3Ddef}), we rewrite $\fnlgrav$ in factorizable form
\be
\fnlgrav(k_1,k_2,k_3) = \frac{1}{12} + \frac{k_1^2}{k_3^2} - \frac{3k_1^4}{2k_3^4} + \frac{3k_1^2k_2^2}{2k_3^4}.
\ee
The resulting CMB bispectrum $\bgrav_{\ell_1\ell_2\ell_3}$ will then be of factorizable form (Eq.~(\ref{eq:factdef})), with $\Nfact=4\Nquad$, 
where $\Nquad$ is the number of quadrature points needed to do the $r$ integral in Eq.~(\ref{eq:WK}).

This example illustrates the ubiquity and power of the factorizability ansatz.
In \cite{Liguori:2005rj}, finding a computationally feasible method for computing Fisher
matrix uncertainties for the gravitational bispectrum at Planck noise levels was left as an unsolved problem.
After representing the bispectrum in factorizable form, we will find, using the algorithms to be
presented in the rest of the paper, that in addition to computing the Fisher matrix rapidly, we can
compute an optimal estimator and construct non-Gaussian simulated maps for this bispectrum.

Next, we consider the ``higher derivative model'' from \citep{Babich:2004gb}, which arises from adding the higher derivative
operator $(\nabla\phi)^4/(8\Lambda^4)$ to the inflationary Lagrangian \cite{Creminelli:2003iq,Alishahiha:2004eh}.
\bea
&& \Fhd(k_1,k_2,k_3) = \frac{9}{7} \fnlhd \Delta_\Phi^2 \Big[ -k_1^{-3} k_2^{-3} \\
&& \qthree + 8 \frac{k_1^{-3}k_2^{-1} + k_1^{-2} k_2^{-1} k_3^{-1}}{(k_1+k_2+k_3)^2} \Big] \symm  \nn
\label{eq:Fhd}
\eea
This expression assumes scale invariance.  In the general case, the amplitude of the three point 
function depends on the dynamics of the field and the expansion of the Universe at the time the triangle of interest 
crosses the horizon during inflation.  For equilateral configurations all three modes in the triangle cross the horizon 
at approximately the same time.

Following a standard convention, we have introduced a parameter $\fnlhd$ for the amplitude of the bispectrum,
normalized so that with all three momenta equal, $F(k,k,k) = 6\fnlhd (\Delta_\Phi/k^3)^2$.
The value of $\fnlhd$ is given by $\fnlhd = (35/432) {\dot\phi}^2/\Lambda^4$, where $\dot\phi$ is the velocity of the inflaton.
(The same bispectrum also arises, typically with a larger value of $\fnlhd$, for DBI inflation \cite{Alishahiha:2004eh}.)

The factor $1/(k_1+k_2+k_3)^2$ appears to ruin factorizability; however, this disease can be cured by introducing a
Schwinger parameter $t$ and writing
\be
\frac{1}{(k_1+k_2+k_3)^2} = \int_0^\infty dt\,\, t e^{-t(k_1+k_2+k_3)}
\label{eq:Sch}
\ee
Using Eqs.~(\ref{eq:Sch}),~(\ref{eq:WK}), one arrives at a CMB bispectrum of the form:
\bea
&& \bhd_{\ell_1\ell_2\ell_3} = \frac{9}{7}\fnlhd \Big[ -\int_0^\infty dr\, r^2 \beta_{\ell_1}(r) \beta_{\ell_2}(r) \alpha_{\ell_3}(r)  \label{eq:bhd}  \\
  && \qquad + 8 \Delta_\Phi^{2/3} \int_0^\infty dt\,t \int_0^\infty dr\,r^2 \Big( \beta_{\ell_1}(r,t) \gamma_{\ell_2}(r,t) \alpha_{\ell_3}(r,t)  \nn \\
  && \qfour + \delta_{\ell_1}(r,t) \gamma_{\ell_2}(r,t) \gamma_{\ell_3}(r,t) \Big) \Big] \symm \nn
\eea
where we have defined:
\bea
\alpha_\ell(r,t)  &=& \frac{2}{\pi} \int_0^\infty dk\,k^2 j_\ell(kr) \Delta^T_\ell(k) e^{-tk}  \\
\beta_\ell(r,t)   &=& \frac{2}{\pi} \int_0^\infty dk\,k^2 j_\ell(kr) \Delta^T_\ell(k) e^{-tk} \left( \frac{\Delta_\Phi}{k^3} \right)  \nn \\
\gamma_\ell(r,t)   &=& \frac{2}{\pi} \int_0^\infty dk\,k^2 j_\ell(kr) \Delta^T_\ell(k) e^{-tk} \left( \frac{\Delta_\Phi}{k^3} \right)^{1/3} \nn \\
\delta_\ell(r,t)   &=& \frac{2}{\pi} \int_0^\infty dk\,k^2 j_\ell(kr) \Delta^T_\ell(k) e^{-tk} \left( \frac{\Delta_\Phi}{k^3} \right)^{2/3} \nn
\eea
(This ordering was chosen so that, for $t=0$, these reduce to the functions 
$\alpha_\ell(r), \beta_\ell(r)$ defined in Eq.~(\ref{eq:alphadef}) and the functions
$\gamma_\ell(r)$, $\delta_\ell(r)$ defined in \cite{Creminelli:2005hu}.)
Note that we have assumed scale invariance throughout our treatment of the higher-derivative bispectrum.
It is seen that the bispectrum is of factorizable form (Eq.~(\ref{eq:factdef})), with $\Nfact = N_1 + 2N_2$,
where $N_1$, $N_2$ are the numbers of quadrature points needed to do the single and double integrals in Eq.~(\ref{eq:bhd}).

Finally, following \cite{Creminelli:2005hu}, we introduce the ``equilateral'' bispectrum
\bea
&& \Feq(k_1,k_2,k_3) = \fnleq \left[ - 3 \frac{\Delta_\Phi^2}{k_1^{4-n_s} k_2^{4-n_s}} \right. \label{eq:Feq}  \\
                   && \qtwo  - 2 \frac{\Delta_\Phi^2}{k_1^{2(4-n_s)/3} k_2^{2(4-n_s)/3} k_3^{2(4-n_s)/3}} \nn  \\
                   && \qtwo \left. + 6 \frac{\Delta_\Phi^2}{k_1^{(4-n_s)/3} k_2^{2(4-n_s)/3} k_3^{4-n_s}} \right] \symm \nn
\eea
In contrast to the other bispectra discussed so far, this bispectrum does not arise from a model;
rather it is designed to approximate the higher-derivative bispectrum (Eq.~(\ref{eq:Fhd})) using fewer factorizable terms.
In \cite{Creminelli:2005hu}, it was shown that the two bispectra are so highly correlated that it suffices to work with
$\Feq$ for data analysis purposes.  We will confirm this result in \S\ref{sec:fisherforecasts}.
The equilateral bispectrum is manifestly of factorizable form, with $\Nfact=3\Nquad$, where $\Nquad$ is the number
of quadrature points needed to do the $r$ integral (Eq.~(\ref{eq:WK})).

As the name suggests, for the equilateral bispectrum (Eq.~(\ref{eq:Feq})), most of the signal-to-noise is contributed
by triples $(\ell_1,\ell_2,\ell_3)$ for which the $\ell$'s are of comparable magnitude.
In contrast, for the local bispectrum (Eq.~(\ref{eq:Flocal})), the greatest contribution is from ``squeezed'' triangles
in which $\ell_1 \ll \ell_2,\ell_3$.
This is reflected in the asymptotics of the 3D bispectra in the squeezed limit $k_2=k_3$, $k_1\rightarrow 0$.
The leading behavior of the local bispectrum is $\bigoh(k_1^{-3})$, whereas the higher-derivative (Eq.~(\ref{eq:Fhd})) 
and equilateral bispectra have leading behavior $\bigoh(k_1^{-1})$ arising from a cancellation between terms.
The gravitational bispectrum (Eq.~(\ref{eq:Fgrav})) has the same $\bigoh(k_1^{-3})$ behavior as the local bispectrum.

\section{Fisher matrix}
\label{sec:fisher}

Before discussing inhomogeneous noise, we first consider the simplest possible noise model for a survey:
homogeneous parameterized by noise power spectrum $N_\ell$.
Such a noise model can be used as an approximation when forecasting sensitivity to different bispectra.
In this noise model, the Fisher matrix of bispectra $B_1, B_2, \ldots$ is defined by
\be
F_{\alpha\beta} \eqdef \frac{1}{6} \sum_{\ell_1\ell_2\ell_3} 
           \frac{(B_\alpha)_{\ell_1\ell_2\ell_3} (B_\beta)_{\ell_1\ell_2\ell_3}}{C_{\ell_1}C_{\ell_2}C_{\ell_3}}  \label{eq:Fdef}
\ee
where $C_\ell = C^{TT}_\ell + N_\ell$ is the total signal + noise power spectrum.
(In Eq.~(\ref{eq:Fdef}), we have written the Fisher matrix for an all-sky survey; for partial sky coverage, one makes the
approximation that the Fisher matrix scales as $F_{\alpha\beta}(\fsky) \propto \fsky F_{\alpha\beta}(1)$.)

The bispectrum covariance obtained from the survey is given by the inverse Fisher matrix:
\be
\Cov(B_\alpha,B_\beta) = (F^{-1})_{\alpha\beta}
\ee
(In particular, the marginalized $1\sigma$ error on bispectrum $B_\alpha$ is given by 
$\sigma_{\rm marg}(B_\alpha) = (F^{-1})_{\alpha\alpha}^{1/2}$
while the error with the other bispectra fixed is given by
$\sigma_{\rm fixed}(B_\alpha) = (F_{\alpha\alpha})^{-1/2}$.)
Thus, the Fisher matrix gives a complete forecast for bispectrum sensitivity of a given survey,
including cross-correlation information, under the simplifying assumption of homogeneous noise.

Let us consider the problem of efficient evaluation of the Fisher matrix in Eq.~(\ref{eq:Fdef}),
assuming that the bispectra under consideration satisfy the factorizability condition
from \S\ref{sec:fact}.

A ``brute force'' harmonic-space approach
is to simply perform the sum over all triples $(\ell_1,\ell_2,\ell_3)$ given in Eq.~(\ref{eq:Fdef}),
evaluating the bispectra $(B_\alpha)_{\ell_1\ell_2\ell_3}$ by straightforward use of
Eqs.~(\ref{eq:bdef}),~(\ref{eq:factdef}).
The computational cost of this approach is $\bigoh(\Nfact\ellmax^3)$.
In many cases, we have found that the harmonic-space approach gives reasonable performance 
and allows the Fisher matrix to be computed straightforwardly.

A second approach is based on computing the Fisher matrix in position space rather than harmonic space.
For notational simplicity we will present the method for the case of a single bispectrum (so that the Fisher
matrix reduces to a number $F$) but our method extends straightforwardly to the multi-bispectrum case.
Writing out the Fisher matrix,
\bea
F &=& \frac{1}{6} \sum_{\ell_1\ell_2\ell_3} \frac{(B_{\ell_1\ell_2\ell_3})^2}{C_{\ell_1}C_{\ell_2}C_{\ell_3}}  \\
  &=& \sum_{\ell_1\ell_2\ell_3ij} \frac{(2\ell_1+1)(2\ell_2+1)(2\ell_3+1)}{144\pi} \threej{\ell_1}{\ell_2}{\ell_3}{0}{0}{0}^2   \nn \\
   && \times \frac{X^{(i)}_{\ell_1} Y^{(i)}_{\ell_2} Z^{(i)}_{\ell_3}}{C_{\ell_1}C_{\ell_2}C_{\ell_3}}
                                \Big[ X^{(j)}_{\ell_1} Y^{(j)}_{\ell_2} Z^{(j)}_{\ell_3} \symm \Big]
\eea
we use the identity
\be
\int_{-1}^1 dz\, P_{\ell_1}(z) P_{\ell_2}(z) P_{\ell_3}(z) = 2 \threej{\ell_1}{\ell_2}{\ell_3}{0}{0}{0}^2
\ee
to write $F$ in the form
\be
F = \sum_{i,j=1}^{\Nfact} F_{ij}  \label{eq:Fsum1}
\ee
where we have defined
\be
F_{ij} \eqdef \frac{2\pi^2}{9} \int_{-1}^1 \Big[\zeta_{X^{(i)}X^{(j)}}\zeta_{Y^{(i)}Y^{(j)}}\zeta_{Z^{(i)}Z^{(j)}} 
                                                \perm \Big] \label{eq:Fsum2}
\ee
where $\perm$ denotes the sum of five additional terms obtained by permuting $\{X^{(j)},Y^{(j)},Z^{(j)}\}$ and
\be
\zeta_{XY}(z) \eqdef \sum_\ell \frac{(2\ell+1)}{4\pi} \frac{X_\ell Y_\ell}{C_\ell} P_\ell(z)
\ee
where $P_\ell(z)$ denotes the Legendre polynomial.

To turn this into an algorithm for computing $F$, we note that the integral in Eq.~(\ref{eq:Fsum2}) can be done
exactly, using Gauss-Legendre integration \cite{NR} with $\Nquad=\lfloor 3\ellmax/2\rfloor + 1$ quadrature points,
since the integrand is a polynomial in $z$ whose degree is $\le 3\ellmax$.
We loop over quadrature points $z$, computing the value of each function $\zeta_{XY}(z)$ which appears, and
accumulating the contribution to each $F_{ij}$ from point $z$, before moving on to the next quadrature point.
This gives a position-space algorithm for Fisher matrix evaluation whose computational cost is $\bigoh(\Nfact^2\ellmax^2)$.
As a rough rule of thumb, we have found that this position-space method is faster when $\Nfact \lesssim 2 \ellmax$, and the
$\bigoh(\Nfact\ellmax^3)$ harmonic-space method is faster when $\Nfact \gtrsim 2 \ellmax$, but the constant 
which appears here will depend on the implementation.

We have introduced the matrix $F_{ij}$ as a device for computing the 1-by-1 ``matrix'' $F$, by summing the entries as in
Eq.~(\ref{eq:Fsum1}), but we note that $F_{ij}$ has a direct interpretation as the $\Nfact$-by-$\Nfact$
Fisher matrix of the individual terms in the factorizable bispectrum (Eq.~(\ref{eq:factdef})).
This observation will be important for the optimization algorithm which we now present.

\section{Optimizing $\Nfact$}
\label{sec:optnfact}

From the preceding discussion, it may seem that our position-space method for Fisher matrix evaluation
is of limited usefulness, providing significant speedup over the harmonic-space method only in the
regime $\Nfact \ll \ellmax$.
However, as we will see in this section, the position-space method also leads to a means of 
``optimizing'' a bispectrum written as a sum of many factorizable terms:
\be
b_{\ell_1\ell_2\ell_3} = \frac{1}{6} \sum_{i=1}^{\Nfact} X^{(i)}_{\ell_1} Y^{(i)}_{\ell_2} Z^{(i)}_{\ell_3} \symm  \label{eq:bnoopt}
\ee
by approximating $b_{\ell_1\ell_2\ell_3}$ by a factorizable bispectrum with fewer terms.
We present an algorithm which retains a subset (of size $\Nopt$) of the original terms and chooses weights $w_1$, \ldots, $w_{\Nopt}$
such that the bispectrum
\be
b'_{\ell_1\ell_2\ell_3} = \frac{1}{6} \sum_{i=1}^{\Nopt} w_i X^{(i)}_{\ell_1} Y^{(i)}_{\ell_2} Z^{(i)}_{\ell_3} \symm  \label{eq:bopt}
\ee
is a good approximation to $b$.
(In Eq.~(\ref{eq:bopt}) and throughout this section, we have assumed for notational simplicity that the terms in the original 
bispectrum (Eq.~(\ref{eq:bnoopt})) have been reordered so that the terms to be retained are in positions 1, \ldots, $\Nopt$.)

Generally speaking, it is only possible to optimize a bispectrum by exploiting redundancy in the factorizable representation,
such as consecutive terms which are nearly equal, or terms which are small enough to be negligible.
For example, a randomly generated factorizable bispectrum cannot be optimized.
The canonical example where optimization is successful is the case where the bispectrum is given exactly by an integral 
over conformal distance $r$ as in Eq.~(\ref{eq:WK}).
In this case, the input bispectrum (Eq.~(\ref{eq:bnoopt})) could be obtained by oversampling the integral using a large number $\Nfact$ of 
quadrature points in $r$.
The output bispectrum (Eq.~(\ref{eq:bopt})) would represent a more efficient quadrature, specifically tailored to the $r$ dependence of
the bispectrum under consideration, using a smaller number $\Nopt$ of points, with integration weights given by the $w_i$.

For Fisher matrix forecasts, it is often unnecessary to optimize $\Nfact$; as discussed in \S\ref{sec:fisher}, the Fisher matrix
can frequently be computed in harmonic space even if the number of factorizable terms is large.
However, for the analysis and simulation algorithms which will be presented in \S\ref{sec:est}-\S\ref{sec:ngsim},
we will see that optimizing $\Nfact$ as a preliminary step usually makes a large improvement in the cost.
This is the main benefit of the optimization algorithm which we now discuss.

\subsection{Optimization algorithm}
\label{ssec:optalg}

Let us first ask: in what sense is one bispectrum $B'_{\ell_1\ell_2\ell_3}$ a good approximation for another bispectrum
$B_{\ell_1\ell_2\ell_3}$?
Our criterion is based on distinguishability; the approximation is good if the Fisher distance
\be
F(B,B') \eqdef \frac{1}{6} \sum_{\ell_1\ell_2\ell_3} 
               \frac{(B_{\ell_1\ell_2\ell_3} - B'_{\ell_1\ell_2\ell_3})^2}{C_{\ell_1}C_{\ell_2}C_{\ell_3}}
\ee
is small.
Here, $C_\ell$ is a signal + noise power spectrum characterizing the survey under consideration, which
is required as an input to our optimization algorithm.
We usually iterate our algorithm until $F(B,B')$ is of order $10^{-6}$ or smaller; this corresponds to
an optimized bispectrum which cannot be distinguished from the original to better than $0.001\sigma$.
(In a realistic survey, the noise will be inhomogeneous and hence not describable by a power spectrum, but
because our termination criterion is so conservative, it suffices to use a noise power spectrum which roughly
models the survey.)

As a first step toward an optimization algorithm, suppose that we have already chosen a subset of terms to retain,
and want to choose optimal (in the sense that the Fisher distance $F(B,B')$ is minimized) values for the weights $w_i$.
In \S\ref{sec:fisher}, we showed (Eq.~(\ref{eq:Fsum2})) how to efficiently calculate the $\Nfact$-by-$\Nfact$ Fisher 
matrix $F_{ij}$ between the individual terms in the input bispectrum (Eq.~(\ref{eq:bnoopt})).
If $F$ is block decomposed into submatrices of size $\Nopt$ and $(\Nfact-\Nopt)$,
\be
F = \left( \begin{array}{cc}
  F_{00}    &  F_{01}  \\
  F_{01}^T  &  F_{11}
\end{array} \right)   \label{eq:Fdecomp}
\ee
then the Fisher distance is given by
\bea
F(B,B') &=& \sum_{ij} (1-w_i) (F_{00})_{ij} (1-w_j)  \label{eq:fbbdef}  \\
         && \qquad + 2 \sum_{iJ} (1-w_i)(F_{01})_{iJ} + \sum_{IJ} (F_{11})_{IJ}  \nn
\eea
Note that we use lower case to denote indices in the first blocks of the decomposition in Eq.~(\ref{eq:Fdecomp}) and upper
case for the second block. The Fisher distance $F(B,B')$ is minimized by choosing
\be
(w_i)_{\rm opt} = 1 + \sum_J (F_{00}^{-1} F_{01})_{iJ}
\ee
and the value at the minimum is given by
\be
F(B,B')_{\rm opt} = \sum_{IJ} (F_{11} - F_{01}^T F_{00}^{-1} F_{01})_{IJ}   \label{eq:fishdist}
\ee

Now that we have seen how to optimize the weights, we address the problem of choosing which terms in the 
original bispectrum (Eq.~(\ref{eq:bnoopt})) to retain.
Mathematically, this corresponds to choosing a subset of terms such that $F(B,B')_{\rm opt}$ (Eq.~(\ref{eq:fishdist})) is minimized.
(Note that after setting the $w_i$ to their optimal values, $F(B,B')_{\rm opt}$ still depends on the subset of terms which are retained, but
this dependence is hidden in Eq.~(\ref{eq:fishdist}), which follows our convention of assuming that the terms have been permuted so that
terms $\{ 1, \ldots, \Nopt\}$ are retained.)
Since exhaustive search of all subsets would be prohibitively slow, our approach is to use a greedy
algorithm: we build up a subset of terms iteratively, in each iteration adding the term which results
in the greatest improvement in $F(B,B')_{\rm opt}$.
The algorithm terminates when $F(B,B')_{\rm opt}$ has reached an acceptably small value.

The simplest implementation of this algorithm would evaluate $F(B,B')_{\rm opt}$ from scratch, using Eq.~(\ref{eq:fishdist}),
for each candidate term in each iteration, which would lead to a running time of $\bigoh(\Nopt^2\Nfact^3)$,
in addition to the time needed to precompute the matrix $F_{ij}$.
This can be improved by two optimizations.
First, in the $n$-th iteration of the algorithm, suppose we have chosen $n$ terms to retain and want
to choose the ($n$+1)-st.
If the matrix $F$ is decomposed into blocks of size $n$ and $(\Nfact-n)$ as in Eq.~(\ref{eq:Fdecomp}), then
one can show that the change in $F(B,B')_{\rm opt}$ if the $I$-th term is added is given by:
\be
\Delta F(B,B')_{\rm opt} = \frac{\left(\sum_J (F_{11} - F_{01}^T F_{00}^{-1} F_{01})_{IJ}\right)^2}
                                {(F_{11} - F_{01}^T F_{00}^{-1} F_{01})_{II}}   \label{eq:deltaF}
\ee
After computing the matrix $(F_{11} - F_{01}^T F_{00}^{-1} F_{01})$, Eq.~(\ref{eq:deltaF}) allows us to select the ($n$+1)-st term
to be retained in time $\bigoh(\Nfact^2)$.
This optimization improves the cost from $\bigoh(\Nopt^2\Nfact^3)$ to $\bigoh(\Nopt\Nfact^3)$;
the limiting step is recomputing the matrix $(F_{11} - F_{01}^T F_{00}^{-1} F_{01})$ from scratch for each $n$.

This brings us to the second optimization: instead of keeping the matrix $F$, in each iteration
we keep the matrix $G$ defined by:
\be
G \eqdef \left( \begin{array}{cc} -F_{00}^{-1} & -F_{00}^{-1}F_{01} \\
                         -F_{01}^T F_{00}^{-1} & F_{11} - F_{01}^T F_{00}^{-1} F_{01} \end{array} \right)
\ee
Note that the lower right block of $G$ is the matrix needed to efficiently select the next term, as described in the previous paragraph.
The other blocks have been constructed so that it is possible to update $G$ in $\bigoh(\Nfact^2)$
time when advancing from the $n$-th iteration of the algorithm to the ($n$+1)-st iteration (rather than
recomputing from scratch at cost $\bigoh(\Nfact^3)$).
More precisely, assuming that terms have been permuted so that the new term to be retained is in the ($n$+1)-st
position, the update rule is given as follows.  If $G$ is given by the block decomposition (into sizes
$n$ and $(\Nfact-n)$)
\be
G = \left( \begin{array}{cc}
   G_{00}    &  G_{01}  \\
   G_{01}^T  &  G_{11}  \\
\end{array} \right) 
= \left( \begin{array}{c|cc}
   G_{00}    &   v_0      &   A_0    \\   \hline
    v_0^T    &  \gamma    &  v_1^T   \\
    A_0^T    &   v_1      &   A_1
\end{array} \right)  \label{eq:Gbldec}
\ee
in the $n$-th iteration, then it is given by
\be
G \rightarrow \left( \begin{array}{cc|c}
G_{00}-(1/\gamma)v_0v_0^T  &  -v_0/\gamma  &  A_0 - (1/\gamma)v_0v_1^T    \\
  -v_0^T/\gamma            &   -1/\gamma   &   -v_1^T/\gamma              \\  \hline
A_0^T - (1/\gamma)v_1v_0^T &  -v_1/\gamma  &  A_1 - (1/\gamma)v_1v_1^T
\end{array} \right)  \label{eq:Gupdate}
\ee
in the $(n+1)$-st iteration.  (Note that the middle ``block'' in Eqs.~(\ref{eq:Gbldec}),~(\ref{eq:Gupdate})
has size 1, e.g. $\gamma$ is a number.)
By keeping the matrix $G$, and using the update rule (Eq.~(\ref{eq:Gupdate})), the cost of the optimization
algorithm is further improved to $\bigoh(\Nopt\Nfact^2)$.

Putting this all together, our optimization algorithm can be summarized as follows.
We initialize the $G$-matrix to the matrix $F_{ij}$ defined in Eq.~(\ref{eq:Fsum2}), 
and initialize the ``score'' $F(B,B')$ to the sum of the entries of $F$.
We then iterate the following loop (starting with $n=0$):
\begin{enumerate}
\item In the $n$-th iteration, we have already chosen a subset of $n$ terms to retain and permuted
the original terms so that these are in positions 1, \ldots, n.  We have stored the $G$-matrix defined
by the block decomposition in Eq.~(\ref{eq:Gbldec}).
\item Choose the index $I = n+1, \ldots, \Nfact$ which maximizes
\be
-\Delta F(B,B')_{\rm opt} = \frac{\left(\sum_J G_{IJ}\right)^2}{G_{II}}  \label{eq:G2F}
\ee
This represents the $(n+1)$-st term chosen to be retained.
\item Update $F(B,B')_{\rm opt}$ according to Eq.~(\ref{eq:deltaF}),
permute the terms so that the new term is in the $(n+1)$-st position,
and update the $G$-matrix according to the rule in Eq.~(\ref{eq:Gupdate}).
\item If $F(B,B')$ has become acceptably small, then terminate, returning the subset of
terms to be retained and the optimal weights
\be
(w_i)_{\rm opt} = 1 - \sum_J G_{iJ}    \label{eq:G2w}
\ee
Otherwise, continue to the $(n+1)$-st iteration.
\end{enumerate}
The total running time is $\bigoh(\Nopt\Nfact^2)$, which in practice is usually less than the $\bigoh(\ellmax^2\Nfact^2)$
time needed to precompute the matrix $F_{ij}$.

\subsection{Optimization examples}
\label{ssec:optex}

\begin{table}
\begin{center}
\begin{tabular}{|c|c|c|c|c|c|}
\hline &  $\nu$  & $\theta_{\rm FWHM}$   & $\Delta_T$   & $\fsky$ & $\ellmax$ \\  \hline
WMAP3   &  41 GHz   &   $28'$  &  460 $\mu\mbox{K}'$   &  0.77  & 1000  \\
        &  61 GHz   &   $20'$  &  560 $\mu\mbox{K}'$   &  0.77  & 1000  \\
        &  94 GHz   &   $12'$  &  680 $\mu\mbox{K}'$   &  0.77  & 1000  \\  \hline
Planck  &  100 GHz  &  $9.2'$  &  51 $\mu\mbox{K}'$    & 0.8  &  2000 \\
        &  143 GHz  &  $7.1'$  &  43 $\mu\mbox{K}'$    & 0.8  &  2000 \\
        &  217 GHz  &  $5.0'$  &  65 $\mu\mbox{K}'$    & 0.8  &  2000 \\  \hline
\end{tabular}
\end{center}
\caption{Experimental parameters used when making Fisher forecasts for three-year WMAP and Planck,
taken from the WMAP3 public data release and \cite{Albrecht:2006um} respectively.}
\label{tab:noise}
\end{table}

We now give some examples of our algorithm for optimizing $\Nfact$, using bispectra from \S\ref{sec:fact}.
We use noise power spectra which roughly approximate the three-year WMAP and Planck surveys, as given in Tab.~\ref{tab:noise}.
(Note that, because the optimization algorithm is Fisher distance based, a noise power spectrum is one of the inputs.)

\begin{figure}
\centerline{\epsfxsize=3.2truein\epsffile[50 500 320 700]{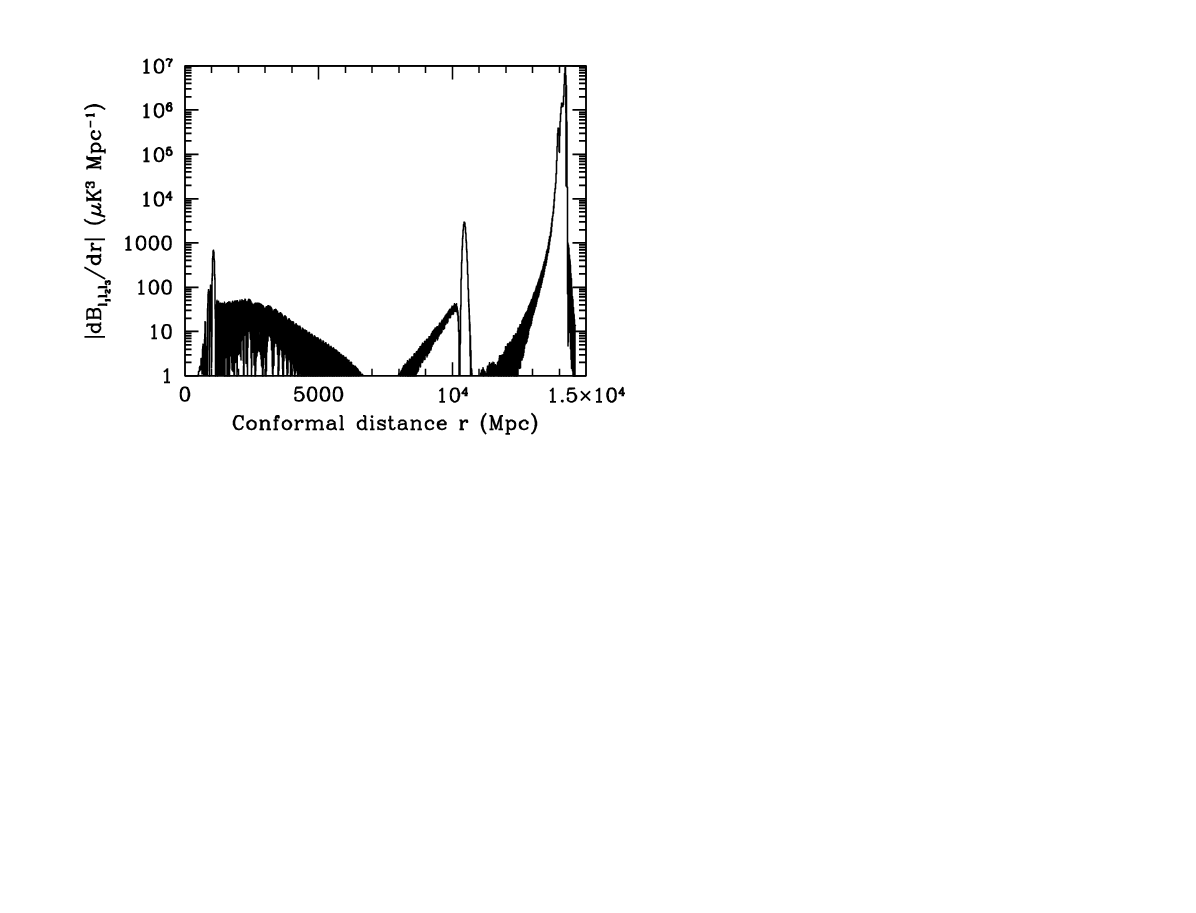}}
\caption{Contribution to $\Bloc_{\ell_1\ell_2\ell_3}$ as a function of conformal distance $r$, with dominant
contribution from recombination ($r \sim 14000$ Mpc), for $(\ell_1,\ell_2\ell_3) = (2,300,300)$, a typical 
squeezed triple with high signal-to-noise.}
\label{fig:fnl_rdependence}
\end{figure}

First consider the local bispectrum $\bloc_{\ell_1\ell_2\ell_3}$, which is an integral over conformal distance $r$ (Eq.~(\ref{eq:blocal})), 
with one factorizable term for each quadrature point needed to do the integral.
To get a feeling for how different values of $r$ contribute, we show the integrand for $(\ell_1,\ell_2,\ell_3)=(2,300,300)$
in Fig.~\ref{fig:fnl_rdependence}.  (This particular triple was selected for high signal-to-noise; the local bispectrum 
is dominated by ``squeezed'' triples with $\ell_1 \ll \ell_2,\ell_3$.)
This structure, showing a large contribution from recombination ($r \sim 14000$ Mpc), with secondary contributions from
reionization ($r \sim 10500$) and ISW ($r \simle 5000$), is typical of bispectra which arise from primordial non-Gaussianity.

\begin{table}
\begin{center}
\begin{tabular}{|c|c|}
\hline $0 \le r \le 9500$       & $\Nquad=150$, linearly-spaced  \\
 $9500 \le r \le 11000$   & $\Nquad=300$, linearly-spaced  \\
 $11000 \le r \le 13800$  & $\Nquad=150$, linearly-spaced  \\
 $13800 \le r \le 14600$  & $\Nquad=400$, linearly-spaced  \\
 $14600 \le r \le 16000$  & $\Nquad=100$, linearly-spaced  \\
 $16000 \le r \le 50000$  & $\Nquad=100$, log-spaced  \\ \hline
\end{tabular}
\end{center}
\caption{Quadrature in $r$ used when computing bispectra in \S\ref{ssec:optex}, with a greater
density of points near reionization (second row) and recombination (fourth row); units for $r$ are Mpc.}
\label{tab:quad}
\end{table}

To be conservative, we oversample the $r$ integral using a quadrature with 
1200 points as shown in Tab.~\ref{tab:quad}.  This quadrature was obtained empirically by
increasing the sampling until each of the bispectra considered in \S\ref{sec:fact} had
converged at the percent level, for several representative choices of $(\ell_1,\ell_2,\ell_3)$.
Using this quadrature for the local bispectrum, we obtain a bispectrum with $\Nfact=1200$.
After running the optimization algorithm, optimized bispectra with $\Nopt=11$ or 21 factorizable terms
are obtained, for WMAP3 or Planck noise levels respectively.
In this case, the optimization algorithm can be thought of as computing a
more efficient quadrature in $r$ by choosing both quadrature points and
integration weights $w_i$.
The resulting quadrature is optimized so that it results in a bispectrum which
is indistinguishable at the given noise levels from the oversampled bispectrum,
while using far fewer quadrature points.

\begin{table}
\begin{center}
\begin{tabular}{|c||c|c|c|}
\hline                                & $\Nfact$  &  $\Nopt$  &  $\Nopt$   \\
                                      & (input)   &  (WMAP3)  &  (Planck)  \\  \hline
Point source (Eq.~(\ref{eq:bps}))     &   1      &    1       &    1   \\
ISW-lensing (Eq.~(\ref{eq:secbispec}))  & 3       &   3       &   3  \\  \hline
Local (Eq.~(\ref{eq:Flocal}))         & 1200      &  11       &  21  \\
Equilateral (Eq.~(\ref{eq:Feq}))      & 3600      &  24       &  47  \\
Gravitational (Eq.~(\ref{eq:Fgrav}))   & 4800      &  168      & 255  \\
HD (Eq.~(\ref{eq:Fhd}))               & 80000     &  33       &  86  \\ \hline
\end{tabular}
\end{center}
\caption{Number of terms $\Nfact$ obtained for the point source, ISW-lensing, local, equilateral, gravitational, and higher derivative bispectra,
after oversampling the integrals using the integration quadratures described in \S\ref{ssec:optex}, and number of terms
$\Nopt$ which are retained after using the optimization algorithm for WMAP3 and Planck noise levels.}
\label{tab:nopt}
\end{table}

Repeating this procedure for the gravitational (Eq.~(\ref{eq:Fgrav})) and equilateral (Eq.~(\ref{eq:Feq})) bispectra, we
arrive at the $\Nopt$ values given in the middle rows of Tab.~\ref{tab:nopt}.
We emphasize that in all cases, the input and output bispectra are instinguishable (to 0.001$\sigma$) at the 
given sensitivity levels, since the optimization algorithm is only allowed to terminate when the Fisher distance
$F(B_{\rm input},B_{\rm opt})$ is $\le 10^{-6}$.

As a final example, consider the higher-derivative bispectrum (Eq.~(\ref{eq:bhd})).
In this case, the bispectrum is a double integral over conformal distance $r$ and Schwinger parameter $t$.
For the $t$ integral, we use a uniform quadrature in $\log(t)$ with three points per decade from $t=10^{-2}$ 
to $t=10^6$.  With this quadrature, we find that the identity
\be
\frac{1}{(k_1+k_2+k_3)^2} = \int_0^\infty dt\,\, t e^{-t(k_1+k_2+k_3)}  \label{eq:Sch2}
\ee
holds to 0.1\%, throughout the range of wavenumbers $10^{-6} \simle k \simle 1$ Mpc$^{-1}$
where the photon transfer function $\Delta_\ell(k)$ is appreciably different from zero
(for $\ellmax=2000$).
As explained in \S\ref{sec:fact}, Eq.~(\ref{eq:Sch2}) is the basis for writing the higher-derivative
bispectrum in factorizable form.
For the $r$ integral, we use the quadrature in Tab.~\ref{tab:quad}, with one additional sublety mentioned
for completeness: at large values of $t$, we have found that the inner integral over $r$ contains 
contributions from large values of $r$.  For this reason, we extend the log-spaced quadrature
in the last row of Tab.~\ref{tab:quad} to $r_{\rm max} = (5 \times 10^6)$ using 400 additional points.

\begin{figure}
\centerline{\epsfxsize=3.2truein\epsffile[50 500 320 700]{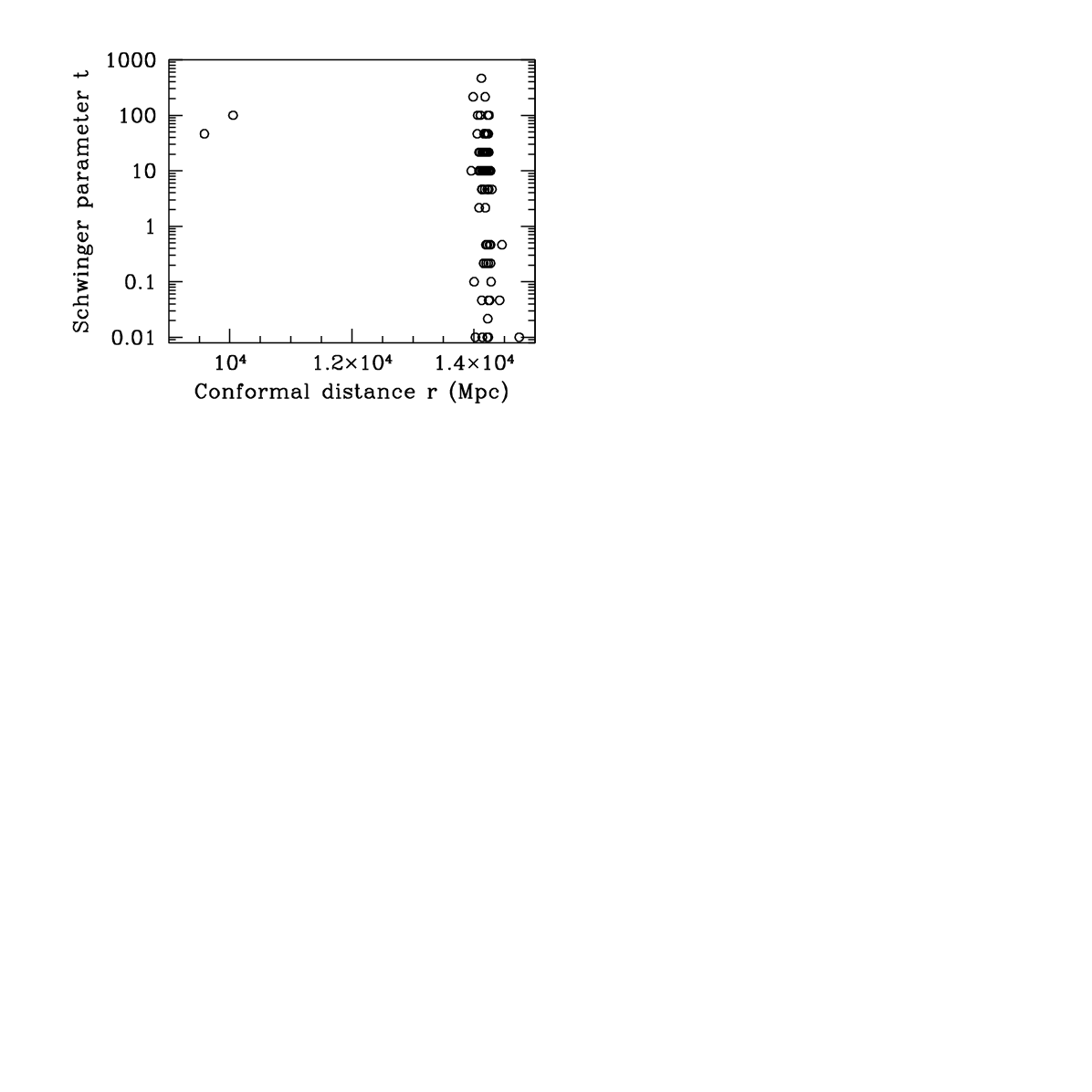}}
\caption{Distribution of factorizable terms in the $(r,t)$ plane for the ``optimized'' higher-derivative
bispectrum $\bhd_{\ell_1\ell_2\ell_3}$.}
\label{fig:trplane}
\end{figure}

Combining this 25-point quadrature in $t$ and 1600-point quadrature in $r$, and with 2 factorizable
terms per quadrature point, we obtain an ``oversampled'' higher-derivative bispectrum with $\Nfact=80000$.
It is infeasible to optimize this bispectrum directly since the cost of the optimization algorithm is
$\bigoh(\Nfact^2\ellmax^2)$.
For this reason, we use a two-level optimization procedure, first separating the factorizable terms into
batches with $\sim 1000$ terms which are optimized separately, then combining the ``optimized batches''
into one bispectrum which is optimized again.
The final result is a bispectrum with $\Nfact=33$ or 86, for WMAP3 or Planck respectively.
For Planck, we show the distribution of terms in the $(r,t)$ plane in Fig.~\ref{fig:trplane}.
Even though the oversampled bispectrum contains terms throughout the ranges $10^{-2} \le t \le 10^6$
and $0 \le r \le (5 \times 10^6)$, the optimization algorithm finds that a much smaller
range in $(r,t)$ suffices to accurately approximate the bispectum.

In more detail, the two-level optimization procedure used for the higher-derivative bispectrum is
given as follows.  The input bispectrum is split into $N$ bispectra $\{B_{\rm input}^{(i)}\}$:
\be
B_{\rm input} = \sum_{i=1}^N B^{(i)}_{\rm input}
\ee
each of which is optimized separately, obtaining bispectra $\{B_{\rm opt}^{(i)}\}$ which satisfy:
\be
F(B_{\rm input}^{(i)}, B_{\rm opt}^{(i)}) \le \frac{10^{-6}}{4 N^2}  \label{eq:twolevel1}
\ee
These are combined for a final round of the optimization algorithm, obtaining an output bispectrum $B_{\rm opt}$
which satisfies:
\be
F\left(B_{\rm opt}, \sum_i B_{\rm opt}^{(i)}\right) \le \frac{10^{-6}}{4}  \label{eq:twolevel2}
\ee
With the threshhold values given on the right-hand sides of Eqs.~(\ref{eq:twolevel1}),~(\ref{eq:twolevel2}), 
it can be proved that the output bispectrum satisfies $F(B_{\rm input}, B_{\rm output}) \le 10^{-6}$, our standard termination criterion.
Thus the accuracy of the approximation need not be degraded by use of the two-level procedure.

An interesting and counterintuitive byproduct of the two-level procedure is that,
even when the number of terms in $B_{\rm input}$ is so large that direct computation of
the 1-by-1 Fisher matrix $F(B_{\rm input}, B_{\rm input})$ is infeasible,
it may be possible to obtain an optimized bispectrum $B_{\rm opt}$ whose Fisher distance 
to $B_{\rm input}$ is provably small.
This increases the scope of the factorizability ansatz: even if the number of terms required
to represent a bispectrum in factorizable form appears to be intractably large, the optimization
procedure may succeed in reducing to a more efficient representation, for which the algorithms
described in this paper will be affordable.

Let us conclude by emphasizing the sense in which the output bispectrum from the optimization algorithm approximates
the input bispectrum.  The only requirement is that the two are experimentally indistinguishable to $0.001\sigma$.
Intuitively, this means that they approximate each other at percent level in regimes which contribute the
most signal-to-noise.  In regimes where the input bispectrum is too small to contribute significant signal-to-noise,
the output bispectrum is not guaranteed to resemble the input; it is only guaranteed also to be small.
This is why, for example, our optimization algorithm tends to drop all contributions after reionization 
($r \simle 9500$); these mainly contribute (via the ISW effect) to triples $(\ell_1,\ell_2,\ell_3)$ in which each 
$\ell_i$ is small, and the total contribution of such triples to the Fisher matrix is negligible.
(In fact, such contributions could presumably be dropped from the outset, but our approach is to conservatively
oversample the integrals and let the optimization algorithm determine which contributions are negligible.)

Our focus will be on Fisher matrix forecasts and bispectrum estimation, for which this distinguishability-based
criterion is ideal, since it permits extremely aggressive optimization of the bispectrum, as seen in Tab.~\ref{tab:nopt}.
However, the optimization algorithm is not without caveats.
As mentioned above, the optimization algorithm is allowed to alter the shape of the bispectrum for triples $(\ell_1,\ell_2,\ell_3)$
in which the bispectrum is so small that it makes a negligible contribution to the Fisher matrix.
One could worry that this could make the optimized bispectrum more sensitive to systematic errors, if there is some observational
reason why these triples are sensitive to systematic effects not modeled by the Fisher formalism.
If this is a concern, then a direct test for systematic contamination can be performed, by verifying that the optimized and
unoptimized shapes give nearly identical values when ``contracted'' with real
data (i.e.~that the values of $T[C^{-1}a]$ are nearly identical, in notation from~\S\ref{sec:est}).
Another concern is that ``closeness'' of the optimized and unoptimized shapes, in the Fisher matrix sense, does not
strictly guarantee that the sampling PDFs of the corresponding three-point estimators must be nearly identical (although it does
guarantee that the two PDFs have variances that are nearly equal).
We verified directly that this is not a concern for the local and equilateral shapes, but defer a more systematic
study of this question for future work.

\section{Forecasts for Planck}
\label{sec:fisherforecasts}

\begin{table}
\begin{center}
\begin{tabular}{|c|cccccc|}
\hline   & Pt. src.  &  ISW     &   Loc.   & Eq.     & Grav.   & HD       \\  \hline
Pt. src. &  (0.05)   &  0.00    &  0.00    & -0.01   & 0.00    & -0.01    \\
ISW      &   0.00    &  (0.16)  &  -0.24   & 0.00    & 0.25    & 0.01     \\
Local    &   0.00    &  -0.24   &  (6.3)   & 0.25    & 0.78    & 0.28     \\
Equil.   &  -0.01    &  0.00    &  0.25    & (66.9)  & 0.36    &  -0.98   \\
Grav.    &   0.00    &  0.25    &  0.77    & 0.36    & (33.4)  & 0.28     \\
HD       &  -0.01    &  0.01    &  0.28    &  -0.98  & 0.28    & (59.9)   \\  \hline
\end{tabular}
\end{center}
\caption{Fisher matrix for Planck, using bispectra from Tab.~\ref{tab:nopt} and noise parameters from Tab.~\ref{tab:noise}.
Off-diagonal entries are correlations; the diagonal parenthesized entries are $1\sigma$ Fisher matrix errors (=$F_{ii}^{-1/2}$) on the
amplitude of each bispectrum, without marginalizing the others.}
\label{tab:planckfisher}
\end{table}

Armed with the optimized bispectra in Tab.~\ref{tab:nopt}, it is straightforward to perform
a Fisher matrix analysis for Planck, using noise parameters from Tab.~\ref{tab:noise}.  The
result is shown in Tab.~\ref{tab:planckfisher}; we obtain $\sigma(\fnlloc)=6.3$, $\sigma(\fnleq)=66.9$
and a $6\sigma$ detection (corresponding to $1\sigma$ error 0.16) of the ISW-lensing bispectrum.
For the point source bispectrum, we have taken a reference value $\bps = 10^{-8}$ $\mu$K$^3$
as described in \S\ref{sec:fact}, and found a $\sim 20\sigma$ detection, but this figure should be
taken very roughly since the value of $\bps$ is very sensitive to the flux limit which is assumed.

One result from the Fisher matrix forecast is that the higher-derivative bispectrum (Eq.~(\ref{eq:Fhd}))
is 98\% correlated to the equilateral shape (Eq.~(\ref{eq:Feq})).  In practice, this means that
it suffices to use the (simpler) equilateral form when analyzing data; the two bispectra cannot
be distinguished (at 1$\sigma$) unless a 25$\sigma$ detection can also be made.  This result agrees well
with \cite{Creminelli:2005hu}, where a 0.98 correlation was also found.

\begin{figure}
\centerline{\epsfxsize=3.2truein\epsffile[50 500 320 700]{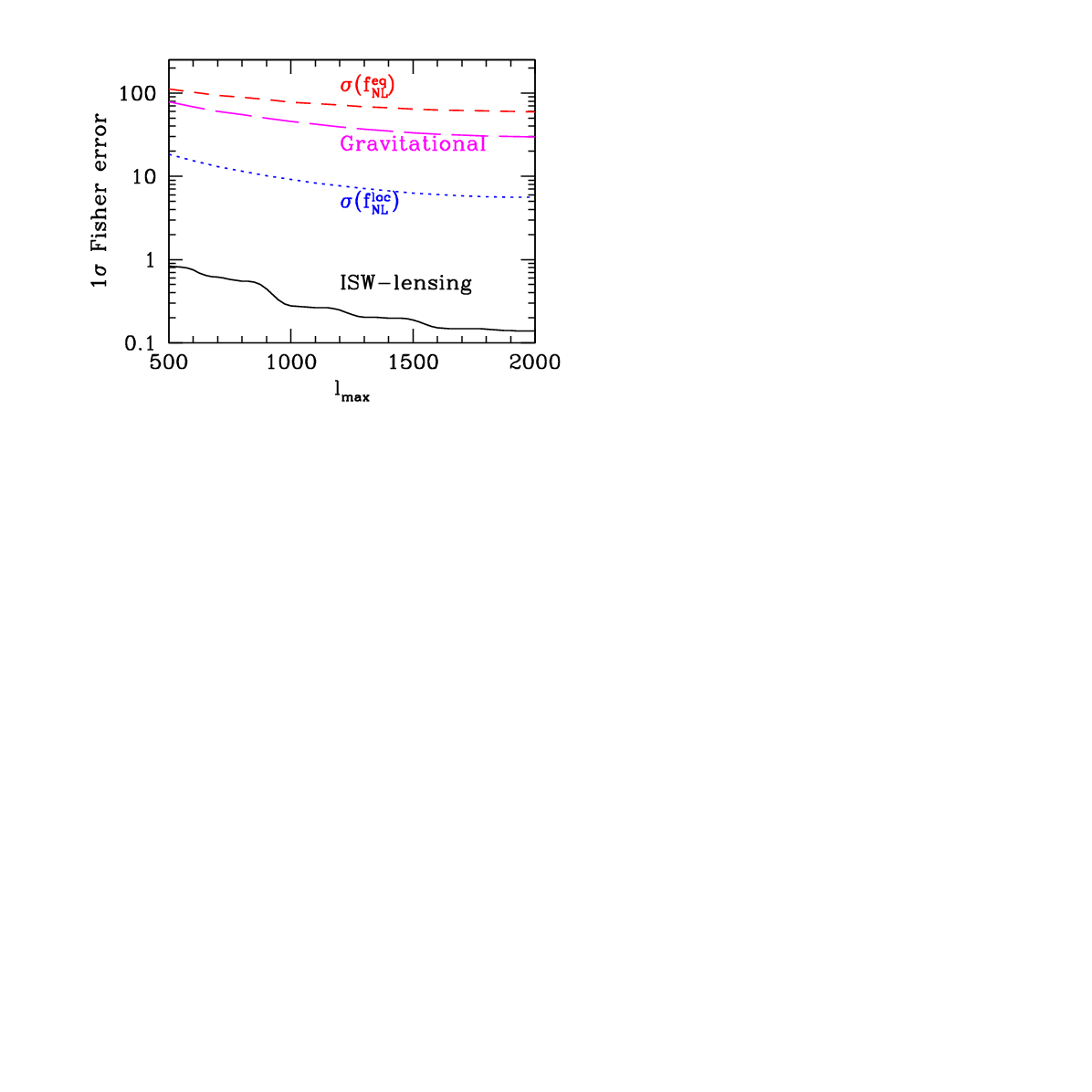}}
\caption{Fisher matrix errors vs. $\ellmax$ for the ISW-lensing, local, equilateral, and gravitational forms of the 
bispectrum, assuming Planck noise levels throughout.}
\label{fig:fisher_lmax}
\end{figure}

A second result is that the gravitational bispectrum (Eq.~(\ref{eq:Fgrav})) is well below the detectability
threshhold for Planck.  (We also find that it is 97\% correlated to a linear combination of the
local and equilateral bispectra, but this is not relevant if its amplitude is too small to be detected.)
One way to quantify this statement is by quoting an ``effective'' value of $f_{NL}$ for which the
local bispectrum (Eq.~(\ref{eq:Flocal})) has the same Fisher matrix error as the gravitational bispectrum (whose
amplitude is assumed fixed).
Using the Fisher matrix we have computed, we obtain $f_{NL}^{\rm eff}=0.2$.
This value does not depend strongly on $\ellmax$, as can be seen in Fig.~\ref{fig:fisher_lmax}, where we show
the dependence of the Fisher matrix errors on $\ellmax$, assuming Planck noise characteristics throughout.

This result seems to disagree with \cite{Liguori:2005rj}, who found $f_{NL}^{\rm eff} \approx 4$ at $\ellmax=500$
with a trend toward increasing $f_{NL}^{\rm eff}$ as $\ellmax$ increases.
Since the two methods for calculating the Fisher matrix are so different, it is difficult to
compare the calculations directly.
However, the result that the gravitational bispectrum is ``weaker'' than the local bispectrum
with $\fnlloc=1$ can be seen intuitively from the 3D bispectra in Eqs.~(\ref{eq:Flocal}),~(\ref{eq:Fgrav}).  One can prove
that the ratio of the two bispectra satisfies
\be
0 \le \left| \frac{\Fgrav(k_1,k_2,k_3)}{\Floc(k_1,k_2,k_3)} \right| \le \frac{13}{12}
\ee
for all values of $(k_1,k_2,k_3)$ which satisfy the triangle inequality,
and the ratio is close to zero for the ``squeezed'' configurations which contribute greatest signal-to-noise.
For example, in the squeezed limit $k_2=k_3$, $k_1\rightarrow 0$, the ratio approaches 1/6.

Another conclusion from the Fisher matrix forecast is that the equilateral bispectrum seems more difficult to detect
than the local bispectrum. This is in some sense just a matter of convention in the definition $f_{NL}$ in the 
different cases.  This has already been observed in the context of WMAP; e.g. in \cite{Creminelli:2006rz}, 
the 1$\sigma$ errors $\sigma(\fnlloc)=34$ and $\sigma(\fnleq)=147$ were obtained from three-year WMAP data.
Here we find (Fig.~\ref{fig:fisher_lmax}) that this trend becomes somewhat more pronounced with increasing $\ellmax$;
for Planck ($\ellmax=2000$), the ratio $\sigma(\fnleq)/\sigma(\fnlloc)$ has increased to 10.6.

\begin{figure}
\centerline{\epsfxsize=1.7truein\epsffile{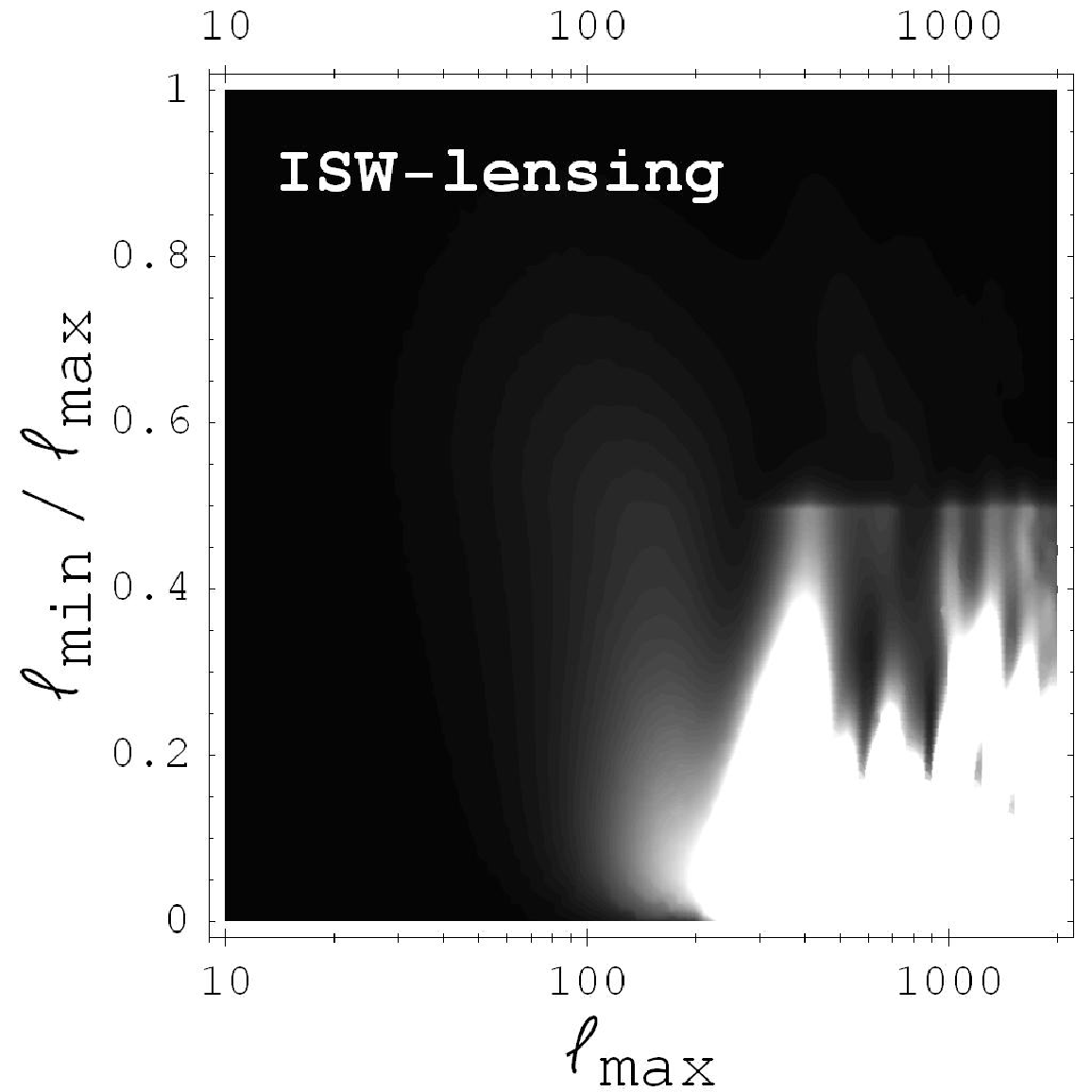} \epsfxsize=1.7truein\epsffile{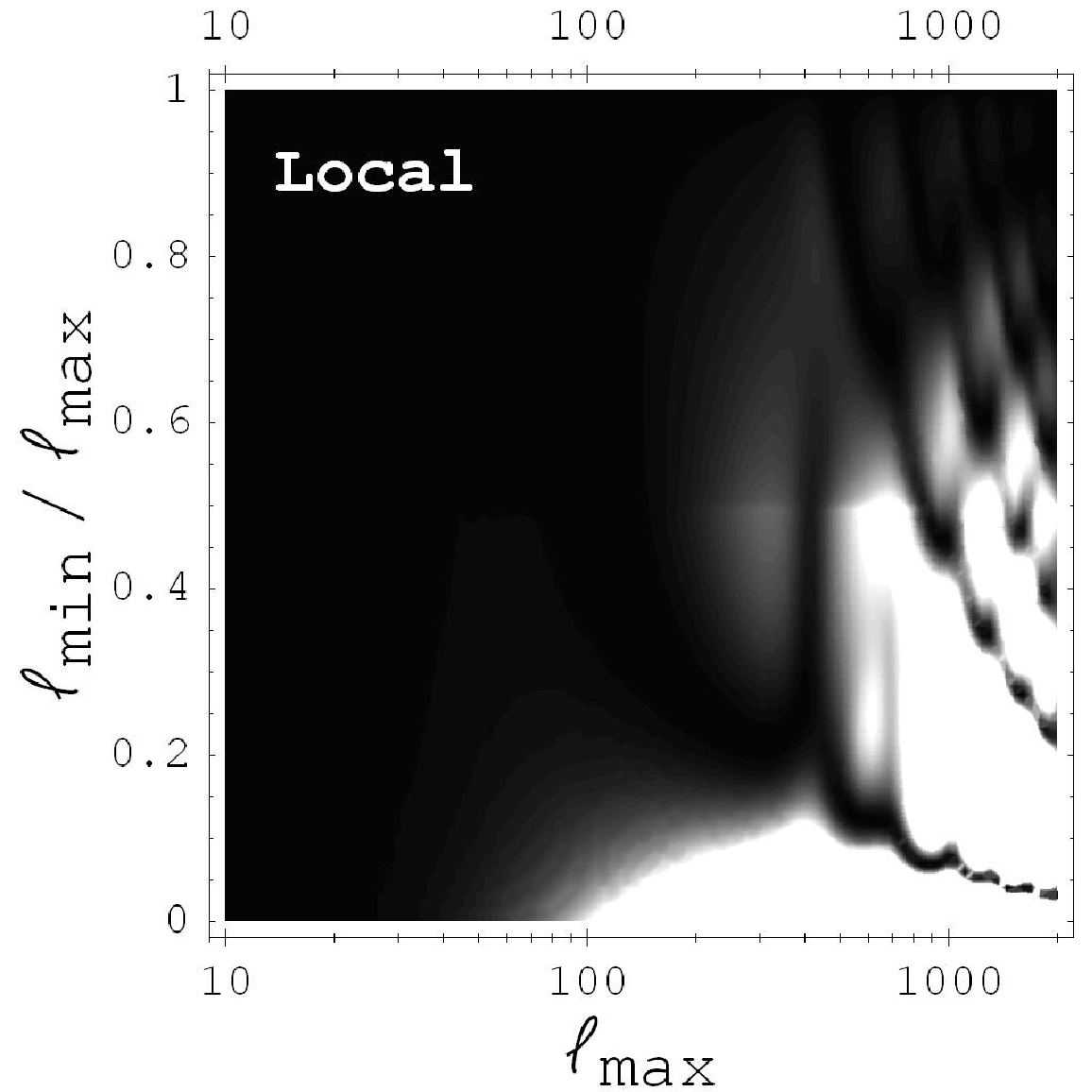}}
\centerline{\epsfxsize=1.7truein\epsffile{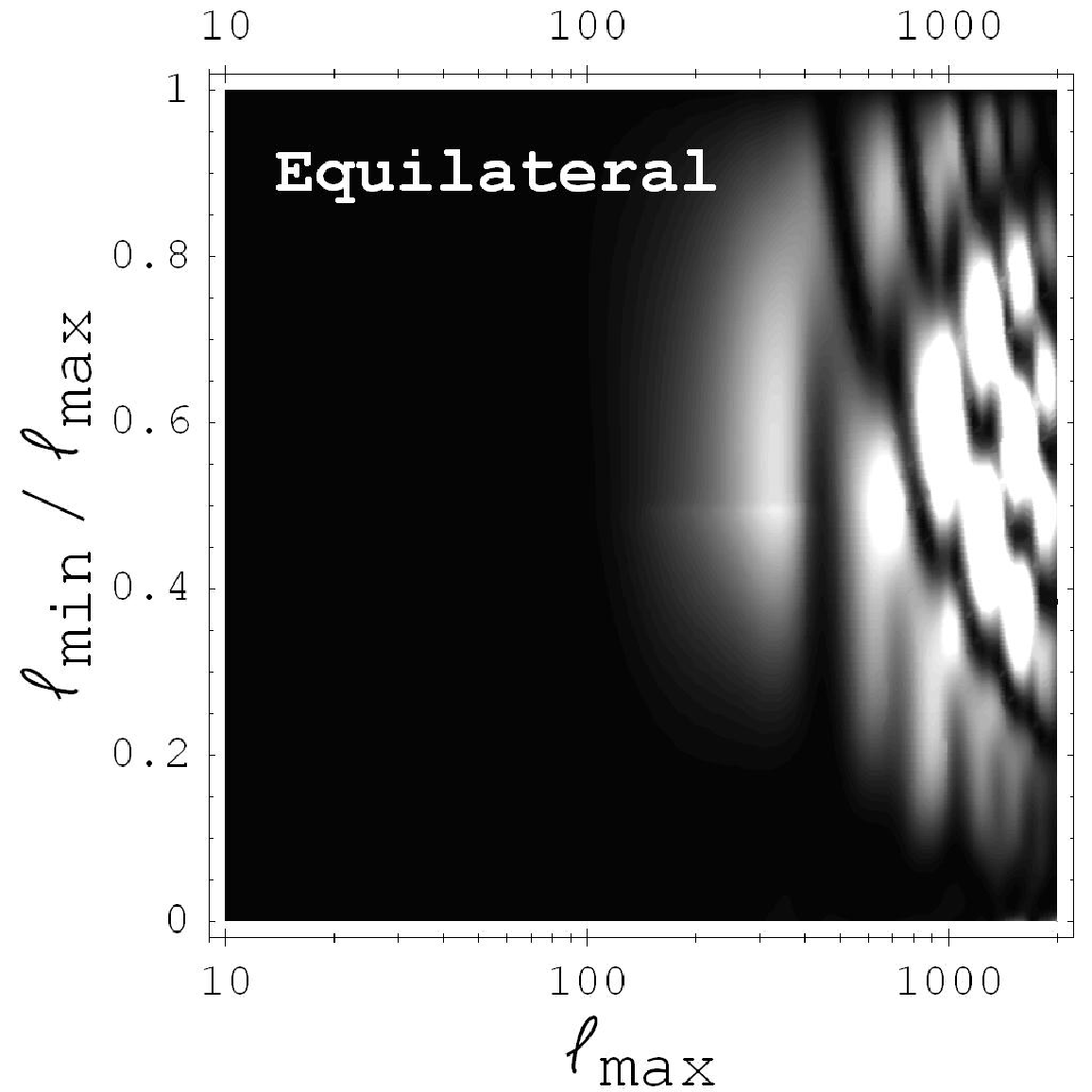} \epsfxsize=1.7truein\epsffile{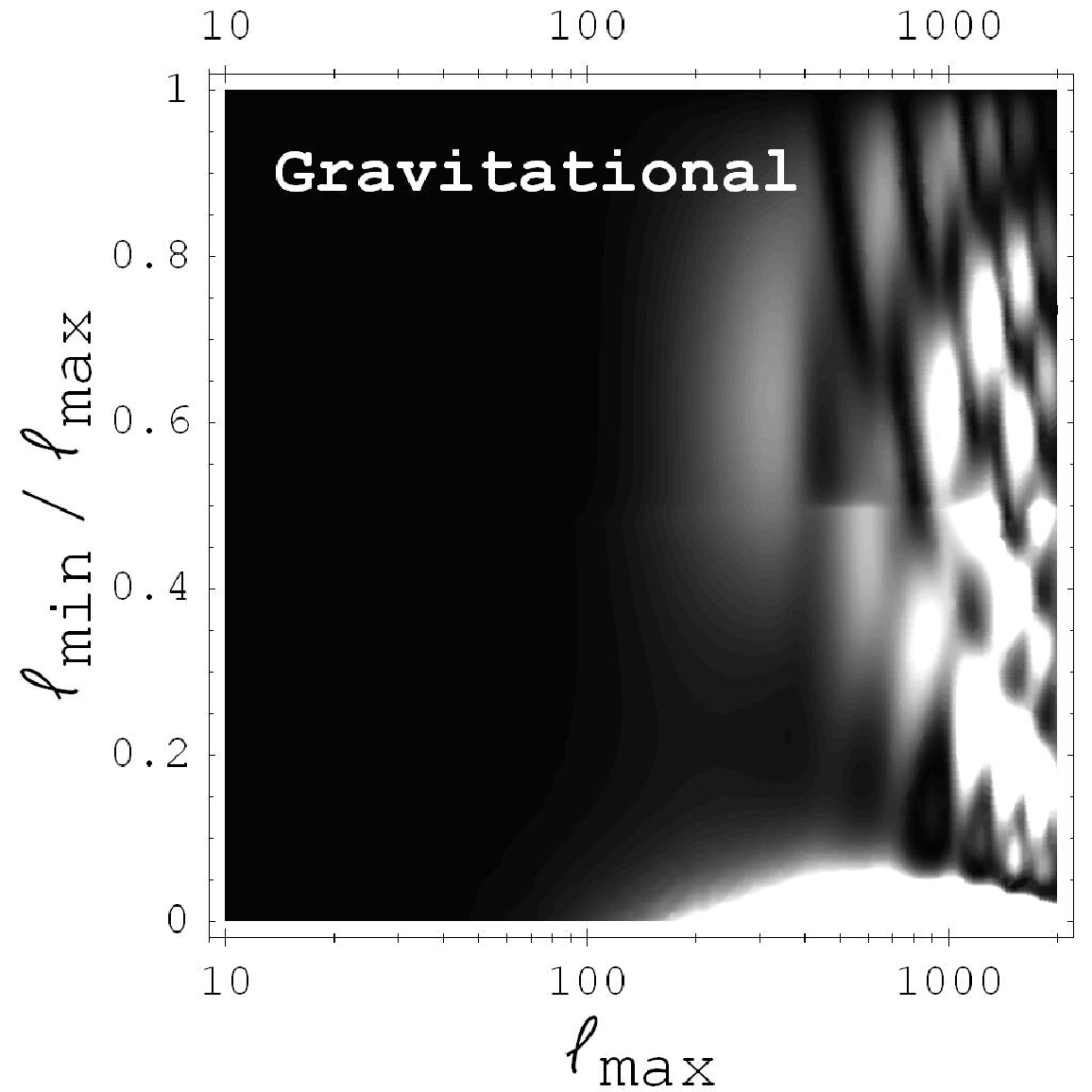}}
\caption{Contour plots of $dF/d(\log\ellmax)d(\ellmin/\ellmax)$, defined in Eq.~(\ref{eq:Fll}),
showing the contribution to the Fisher matrix error as a function of $(\ellmin,\ellmax)$, for the ISW-lensing,
local, equilateral, and gravitational forms of the bispectrum.}
\label{fig:fplot}
\end{figure}

Finally, we find that correlations between the point source, ISW-lensing, local, and equilateral bispectra are small.
In effect, the CMB gives independent constraints on these four bispectra which are not appreciably
degraded by marginalizing the other three.
However, the correlation between the ISW-lensing and local bispectra is large enough that the presence
of the former (a guaranteed $6\sigma$ signal) contributes non-negigible bias ($\Delta\fnlloc = 9.8$) when 
estimating the latter.
If a multi-bispectrum analysis with marginalization is performed, then this bias will be subtracted without
significantly degrading $\sigma(\fnlloc)$.
A similar comment applies to the point source bispectrum; we have found that at the Planck reference value 
($\bps = 10^{-8}$ $\mu$K$^3$), there is negligible bias contributed to the other bispectra, but the point source bispectrum
should be marginalized in practice since its value is quite uncertain.

It is illuminating to show the contributions to the Fisher matrix from differently shaped triples $(\ell_1,\ell_2,\ell_3)$.
In Fig.~\ref{fig:fplot}, we show contour plots of the quantity
\be
\frac{dF}{d\log(\ellmax)\,d(\ellmin/\ellmax)} = \ellmax^2 \sum_{\ell=\ellmin}^{\ellmax} \frac{(B_{\ellmin,\ell,\ellmax})^2}{6 C_{\ellmin} C_\ell C_{\ellmax}}  \label{eq:Fll}
\ee
The Jacobian $(\ellmax^2)$ has been included as a prefactor so that the Fisher matrix element $F$
will be given by integrating over the variables $\{\log(\ellmax), (\ellmin/\ellmax)\}$ on the axes of the plot.
The sharp feature seen in these plots at $\ellmin/\ellmax = (1/2)$ arises solely from the behavior of the
Wigner 3j symbols, and would be present even if the reduced bispectrum $b_{\ell_1\ell_2\ell_3}$ were
simply proportional to $(C_{\ell_1}C_{\ell_2}C_{\ell_3})^{1/2}$.

\begin{figure}
\centerline{\epsfxsize=3.2truein\epsffile[50 500 320 700]{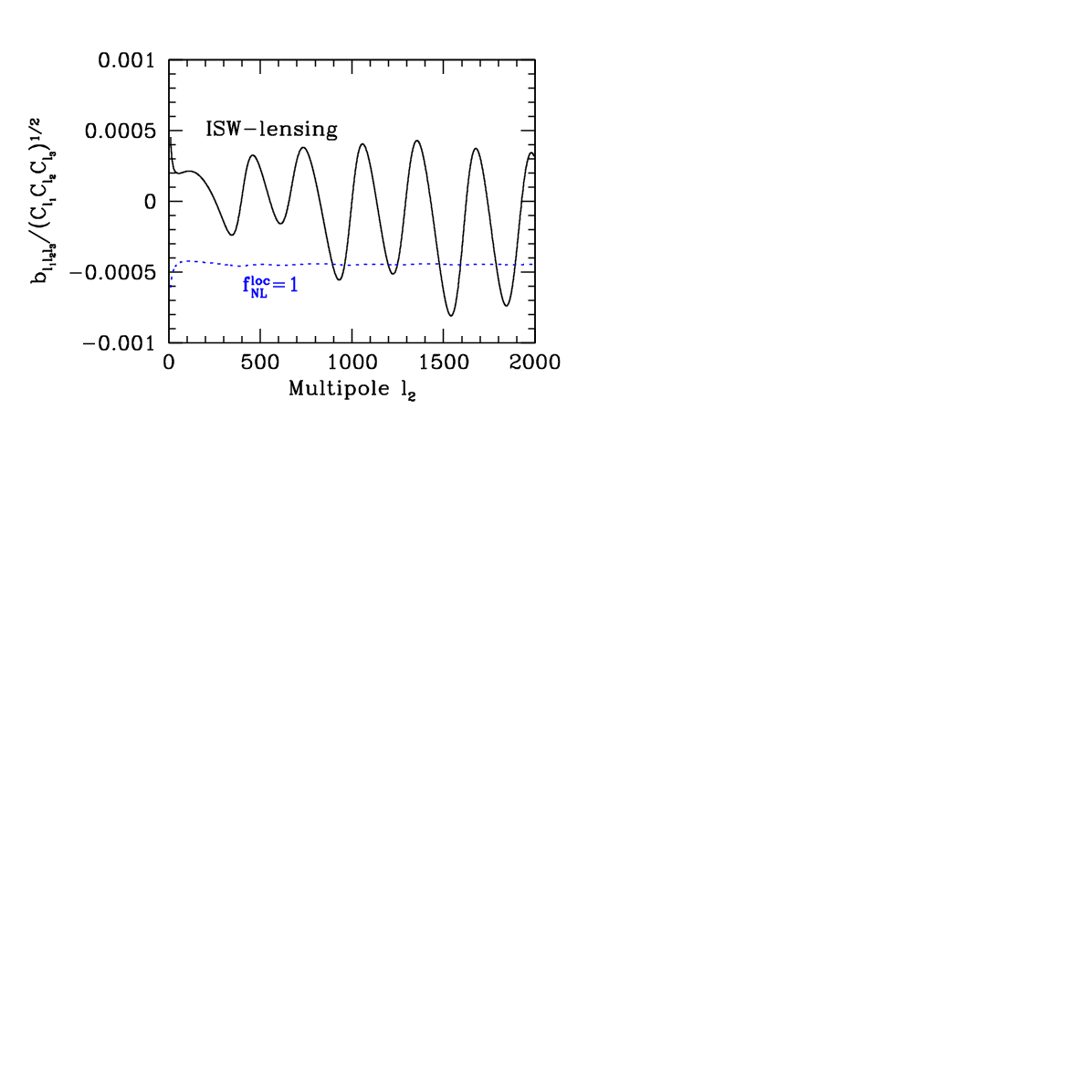}}
\caption{Values of $b_{\ell_1,\ell_2\ell_3}/(C_{\ell_1}C_{\ell_2}C_{\ell_3})^{1/2}$, showing
oscillations in the ISW-lensing bispectrum but not the local bispectrum, plotted for
$\ell_1=10$ and $\ell_3=\ell_2+6$.  This is a typical ``squeezed'' triangle which contributes
high signal-to-noise in both cases.}
\label{fig:osc}
\end{figure}

It is seen that the equilateral bispectrum receives most of its signal-to-noise from the ``equilateral'' regime ($\ellmin$ comparable
to $\ellmax$), whereas the ISW-lensing and local bispectra receive highest contributions from the squeezed regime ($\ellmin \ll \ellmax$),
confirming the intuitive picture from the end of \S\ref{sec:fact}.
From this description it may seem surprising that the correlation between the ISW-lensing and local bispectra is so small (0.25).
This is because the ISW-lensing bispectrum contains oscillatory features which are not present in the local case (Fig.~\ref{fig:osc}).
These are hidden in Fig.~\ref{fig:fplot} and help orthogonalize the two bispectra.
If one were to artificially replace $b_{\ell_1\ell_2\ell_3} \rightarrow |b_{\ell_1\ell_2\ell_3}|$, then the correlation between the
ISW-lensing and local bispectra would increase to 0.6.

\section{Optimal bispectrum estimation}
\label{sec:est}

In this section, we present a general framework for optimal bispectrum estimation
in the presence of sky cuts and inhomogeneous noise.
In the context of power spectrum estimation, a method similar to ours was 
proposed by \citet{Oh:1998sr}, and it is illuminating to first revisit their construction.

\subsection{Power spectrum estimation}
\label{sec:psest}

The power spectrum estimation problem can be stated as follows.  One is interested in
simultaneously estimating the amplitude of bandpowers $C_1, \ldots, C_{\Nband}$ from a
map $m$ with noise covariance $N$.  It can be shown \citep{Bond:1998zw}
that the optimal estimator $\estE_\alpha$ for each bandpower $\alpha$ is given by
\be
\estE_\alpha[m] \eqdef \frac{1}{2} F_{\alpha\beta}^{-1} \big( m^T C^{-1} C_\beta C^{-1} m \big)\,,    \label{eq:ps1}
\ee
where $C=S+N$ is the total signal + noise, $m$ is the observed temperature map (assumed
to have covariance $C$), and $F_{\alpha\beta}$ is the Fisher matrix 
\be
F_{\alpha\beta} \eqdef \frac{1}{2} \Tr\big[ C_\alpha C^{-1} C_\beta C^{-1} \big]\,.    \label{eq:ps2}
\ee
A subtlety in Eqs.~(\ref{eq:ps1}),~(\ref{eq:ps2}) is that, strictly speaking, the definition of the estimator
depends (via the matrix $C^{-1}$) on the signal covariance $S$ which one is trying to estimate in the first place.
In practice, the estimator can be iterated until convergence is reached; it can be shown that the
limiting value of $S$ obtained in this way is equal to a maximum likelihood estimate \citep{Bond:1998zw}.
This subtlety will be ignored in this section where our purpose is merely to set the stage, in the more
familiar context of power spectrum estimation, for the bispectrum estimator which follows.

Evaluating Eqs.~(\ref{eq:ps1}),~(\ref{eq:ps2}) appears infeasible for large maps owing to the $\bigoh(\Npix^3)$ matrix
operations which appear.  However, the computational cost can be reduced by avoiding use of dense matrices
(e.g.~\cite{Oh:1998sr,Borrill:1998pn}).
Considering Eq.~(\ref{eq:ps1}) first, one only needs to multiply a single map by $C^{-1}$, which can
often be done affordably (and without needing to store the matrix $C$ in dense form)
using conjugate gradient inversion, although the details will depend on the
experiment's noise model.
Considering next Eq.~(\ref{eq:ps2}), the trace can be written as as a Monte Carlo average:
\be
F_{\alpha\beta} = \frac{1}{2} \left\langle a^T C^{-1} C_\alpha C^{-1} C_\beta C^{-1} a \right\rangle_a\,,  \label{eq:ohsmc}
\ee
where the notation $\langle\cdot\rangle_a$ denotes an average taken over signal + noise realizations $a$
(i.e., $a$ is a Gaussian random field with covariance $C$).
If we continue to assume an affordable procedure for multipliying a map by $C^{-1}$, Eq.~(\ref{eq:ohsmc}) 
permits $F_{\alpha\beta}$ to be computed by Monte Carlo.
The estimator covariance is then given by
\be
\Cov(\estE_\alpha,\estE_\beta) = F_{\alpha\beta}^{-1}\,.   \label{eq:ps3}
\ee
The matrix $F_{\alpha\beta}$, defined by Eq.~(\ref{eq:ps2}), is the Fisher matrix for the survey with
noise covariance given by an arbitrary matrix $N$.
For optimal estimators, the Fisher matrix gives both the normalization (Eq.~(\ref{eq:ps1})) and 
the covariance (Eq.~(\ref{eq:ps3})).

This method for optimal power spectrum estimation in the presence of arbitrary noise covariance $C$
has an analogue for bispectra, as we will now see.

\subsection{Bispectrum estimation}

Let us now consider the analagous problem of optimal estimation of the amplitude of a given bispectrum $B_{\ell_1\ell_2\ell_3}$.
The form of the optimal estimator has been constructed previously \cite{Babich:2005en} and shown to contain
both cubic and linear terms:
\bea
\estE[a] &=& \frac{1}{6F_{\estE}} B_{\ell_1\ell_2\ell_3} \threej{\ell_1}{\ell_2}{\ell_3}{m_1}{m_2}{m_3} \nn \\
       &\times& \Big[ C^{-1}_{\ell_1 m_1 \ell_4 m_3} C^{-1}_{\ell_2 m_2 \ell_5 m_5}  C^{-1}_{\ell_3 m_3 \ell_6 m_6} a_{\ell_4 m_4} a_{\ell_5 m_5} a_{\ell_6 m_6}  \nn \\
      &&  \qtwo - 3 C^{-1}_{\ell_1 m_1 \ell_2 m_2} C^{-1}_{\ell_3 m_3 \ell_4 m_4} a_{\ell_4 m_4} \Big] \label{eq:Edef}
\eea
where $F_{\estE}$ is a constant which normalizes the estimator to have unit response to $B_{\ell_1\ell_2\ell_3}$.
(The factor 1/6 has been included for later convenience.)
In order to translate the value of $\estE$ to a constraint on the bispectrum amplitude,
one needs to know both the normalization $F_{\estE}$ and the variance $\Var(\estE)$.

The cubic term in the estimator can be thought of as a matched filter whose shape is given by the bispectrum 
$B_{\ell_1\ell_2\ell_3}$.  The linear term can improve the variance of the estimator for certain bispectra (more precisely,
bispectra of ``squeezed'' shape).  For example, better limits on 
$\fnlloc$ are obtained from the one-year WMAP data using an estimator containing the linear term than from the three-year
WMAP data without the linear term \cite{Creminelli:2005hu,Spergel:2006hy}.
We note that for the fully optimal estimator (Eq.~(\ref{eq:Edef})), the data
appears weighted by inverse signal + noise $C^{-1}$, and so the variance of the estimator always improves as more 
modes are added to the data.  
This is in contrast to suboptimal methods, such as those used to analyze WMAP data to date
\cite{Komatsu:2003fd,Creminelli:2005hu,Spergel:2006hy,Creminelli:2006rz},
where as the cutoff multipole $\ellmax$ is increased, the variance eventually worsens as 
a result of the inhomogeneities in the noise and the sky cuts. 

We now introduce notation which will be used throughout the rest of the paper.
Given a map $a = \{a_{\ell m}\}$, define
\be
T[a] \eqdef \frac{1}{6} \sum_{\ell_i m_i}  B_{\ell_1\ell_2\ell_3} \threej{\ell_1}{\ell_2}{\ell_3}{m_1}{m_2}{m_3} 
                               a_{\ell_1 m_1} a_{\ell_2 m_2} a_{\ell_3 m_3}
\label{eq:Tdef}
\ee
We also consider the gradient of $T[a]$ with respect to the input map:
\be
\nabla_{\ell m} T[a] \eqdef \frac{\partial T[a]}{\partial a_{\ell m}^*} = \frac{1}{2} 
                            B_{\ell\ell_2\ell_3} \threej{\ell}{\ell_2}{\ell_3}{m}{m_2}{m_3} a_{\ell_2 m_2}^* a_{\ell_3 m_3}^*  \label{eq:gradT}
\ee
Note that $T[a]$ is a scalar which is cubic in the input map $a$, 
whereas $\nabla T[a]$ is another map which is quadratic in $a$.

The significance of $T$ is that three quantities of interest can be written compactly as Monte Carlo
averages involving $T$, $\nabla T$.
First, the estimator (Eq.~(\ref{eq:Edef})) can be rewritten:
\bea
\estE[a] &=& \frac{1}{F_{\estE}} \Big( T[C^{-1}a] -  \label{eq:mcE} \\
&& a_{\ell_1 m_1} C^{-1}_{\ell_1 m_1,\ell_2 m_2} \Big\langle \nabla_{\ell_2 m_2} T[C^{-1} a'] \Big\rangle_{a'} \Big) \nn
\eea
obtaining the linear term as a Monte Carlo average.
Here, $\langle \cdot \rangle_{a'}$ denotes the Monte Carlo average taken over Gaussian realizations $a'$ of signal + noise.
Second, the normalization constant is given by:
\bea
F_{\estE} &=& \frac{1}{3} \Big\langle \nabla_{\ell_1 m_1} T[C^{-1}a] C^{-1}_{\ell_1 m_1,\ell_2 m_2} \nabla_{\ell_2 m_2} T[C^{-1}a] \Big\rangle_a  \label{eq:mcF} \\
           &&   - \frac{1}{3} \Big\langle \nabla_{\ell_1 m_1} T[C^{-1}a] \Big\rangle_a 
                               C^{-1}_{\ell_1 m_1,\ell_2 m_2} \Big\langle \nabla_{\ell_2 m_2} T[C^{-1}a] \Big\rangle_a  \nn 
\eea
Third, the estimator variance is given by
\be
\Var(\estE) = F_{\estE}^{-1}.
\label{eq:estvar}
\ee
Eqs.~(\ref{eq:Edef})-(\ref{eq:estvar}) are derived in Appendix \ref{app:mc}.
Taken together, they provide the basis for the following Monte Carlo procedure, which is the main result of this section:

\begin{enumerate}
\item In each Monte Carlo iteration, construct a random signal + noise realization $a$.
\item Evaluate $C^{-1}a$, $\nabla T[C^{-1}a]$, and $C^{-1}(\nabla T[C^{-1}a])$ (see below).
\item Accumulate the contribution to the following Monte Carlo averages:
\be
 \Big\langle \nabla_{\ell_1 m_1} T[C^{-1}a] \Big\rangle_a 
\ee
\be
 \Big\langle \nabla_{\ell_1 m_1} T[C^{-1}a] C^{-1}_{\ell_1 m_1, \ell_2 m_2} \nabla_{\ell_2 m_2} T[C^{-1}a] \Big\rangle_a  \\
\ee
\end{enumerate}
At the end of the Monte Carlo loop, the linear term in the estimator (Eq.~(\ref{eq:mcE})),
the estimator normalization (Eq.~(\ref{eq:mcF})), and the estimator variance (Eq.~(\ref{eq:estvar}))
have been computed.
For each of these three quantities, the Monte Carlos converge to a value which incorporates
the full noise covariance (including the sky cut), in contrast to methods which assume an
$\fsky$ scaling.
A feature of the method is that the variance (Eq.~(\ref{eq:estvar})) includes the effect of the linear term in the estimator,
even though the linear term is not precomputed.  Both are computed in parallel by the same set of Monte Carlos.
After the Monte Carlo loop, one can evaluate the estimator (in the form given by Eq.~(\ref{eq:mcE})),
including linear term and normalization, on the observed CMB map $m$.

In each Monte Carlo iteration, one $\nabla T$ evaluation and two multiplications by $C^{-1}$
are required.  For the first of these ingredients, evaluating $\nabla T$, we will give a fast algorithm 
in the next section, assuming that the bispectrum satisfies the factorizability condition (Eq.~(\ref{eq:factdef})).

A method for the second ingredient, multiplying a map by $(S+N)^{-1}$, will depend on the noise model
of the experiment under consideration.
In this paper, where our emphasis is on general algorithms which are not experiment-specific, we will not
address this problem.
However, let us emphasize that the experiment-specific challenge of finding an affordable method for $(S+N)^{-1}$
multiplication is a necessary ingredient in the {\em optimal} estimators of Eq.~(\ref{eq:ps1}) and  Eq.~(\ref{eq:Edef}).
If this problem can be solved, then the general framework we have presented here will permit optimal bispectrum estimation.
Otherwise, one must fall back on a suboptimal estimator, e.g. replacing $C^{-1}$ in Eq.~(\ref{eq:Edef}) by 
a filter which approximates it.
Since it may not be feasible to solve the $(S+N)^{-1}$ problem for every dataset in which the three-point function
is studied, we treat this case in App.~\ref{app:subopt}, including a discussion of how the linear term which improves
the estimator variance may be retained, even when full optimality is lost.

We conclude this section by describing the generalization of our Monte Carlo procedure to
joint estimation of multiple bispectra $B_1, \cdots, B_n$.
Denote the quantity $T[a]$ (Eq.~(\ref{eq:Tdef})), evaluated using bispectrum $B_\alpha$, by $T_\alpha[a]$.
Then the optimal estimator for $B_\alpha$, debiased to have zero mean response to $B_\beta$ ($\beta\ne \alpha$), 
is given by:
\be
\estE_\alpha = F_{\alpha\beta}^{-1} \left( T_\beta[C^{-1}a] - a_{\ell_1 m_1} C^{-1}_{\ell_2 m_2} 
                                                \Big\langle \nabla_{\ell_2 m_2} T_\beta[C^{-1} a'] \Big\rangle_{a'} \right)  \label{eq:bFdef}
\ee
where $F_{\alpha\beta}$ is the matrix
\bea
  && F_{\alpha\beta} = \frac{1}{3} \Big\langle \nabla_{\ell_1 m_1} T_\alpha[C^{-1}a] C^{-1}_{\ell_1 m_1, \ell_2 m_2} 
                                     \nabla_{\ell_2 m_2} T_\beta[C^{-1}a] \Big\rangle_a   \\
        && \quad  - \frac{1}{3} \Big\langle \nabla_{\ell_1 m_1} T_\alpha[C^{-1}a] \Big\rangle_a  \label{eq:Fabdef}
                          C^{-1}_{\ell_1 m_1, \ell_2 m_2} \Big\langle \nabla_{\ell_2 m_2} T_\beta[C^{-1}a] \Big\rangle_a  \nn
\eea
The estimator covariance is given by
\be
\Cov(\estE_\alpha,\estE_\beta) = F_{\alpha\beta}^{-1}\,.  \label{eq:bFcov}
\ee
With multiple bispectra, the Monte Carlo procedure is the same as described above.  In each iteration,
$C^{-1}a$, $\nabla T_\alpha[C^{-1}a]$, and $C^{-1}\nabla T_\alpha[C^{-1}a]$ are computed, for a total of $n$ evaluations
of $\nabla T$ and $(n+1)$ multiplications by $C^{-1}$.

The matrix $F_{\alpha\beta}$ defined in Eq.~(\ref{eq:Fabdef}) is the $n$-by-$n$ Fisher matrix between the bispectra $B_1, \cdots, B_n$,
with the full noise properties of the survey incorporated via the covariance $C$.
(In the case where the noise is homogeneous, $F_{\alpha\beta}$ reduces to the Fisher matrix that was defined previously in Eq.~(\ref{eq:Fdef})).
In the estimator framework, $F_{\alpha\beta}$ arises as both the normalization (Eq.~(\ref{eq:bFdef}))
and the covariance (Eq.~(\ref{eq:bFcov})) of optimal bispectrum estimators.
This is in complete analogy with optimal power spectrum estimation as described in \S\ref{sec:psest}.

\subsection{Comparison with direct MC}
\label{ssec:directmc}

We have obtained the estimator covariance and normalization (which are equal for the case of
the optimal estimator in Eq.~(\ref{eq:Edef})) as a Monte Carlo average (Eq.~(\ref{eq:mcF})).
The reader may be wondering why a simpler Monte Carlo procedure was not used instead.
For example, consider the variance of the cubic term $T[C^{-1}a]$ in the estimator.
Following the treatment above, this would be computed via the Monte Carlo prescription
\bea
&& \Var(T[C^{-1}a]) = \label{eq:mcfancy} \\
&& \qquad \frac{1}{3} \Big\langle \nabla_{\ell_1 m_1} T[C^{-1}a] 
                   C^{-1}_{\ell_1 m_1,\ell_2 m_2} \nabla_{\ell_2 m_2} T[C^{-1}a] \Big\rangle_a  \nn \\
&& \qquad  + \frac{2}{3} \Big\langle \nabla_{\ell_1 m_1} T[C^{-1}a] \Big\rangle_a  
                               C^{-1}_{\ell_1 m_1,\ell_2 m_2} \Big\langle \nabla_{\ell_2 m_2} T[C^{-1}a] \Big\rangle_a  \nn
\eea
(Compare Eq.~(\ref{eq:mcF}), which gives the variance when the linear term is included in addition to the cubic term.)
Why not compute this more simply by using the ``direct'' Monte Carlo prescription
\be
\Var(T[C^{-1}a]) = \langle T[C^{-1}a] T[C^{-1}a] \rangle_a   \label{eq:mcdirect}
\ee
instead?

\begin{figure}
\centerline{\epsfxsize=3.2truein\epsffile[50 500 320 700]{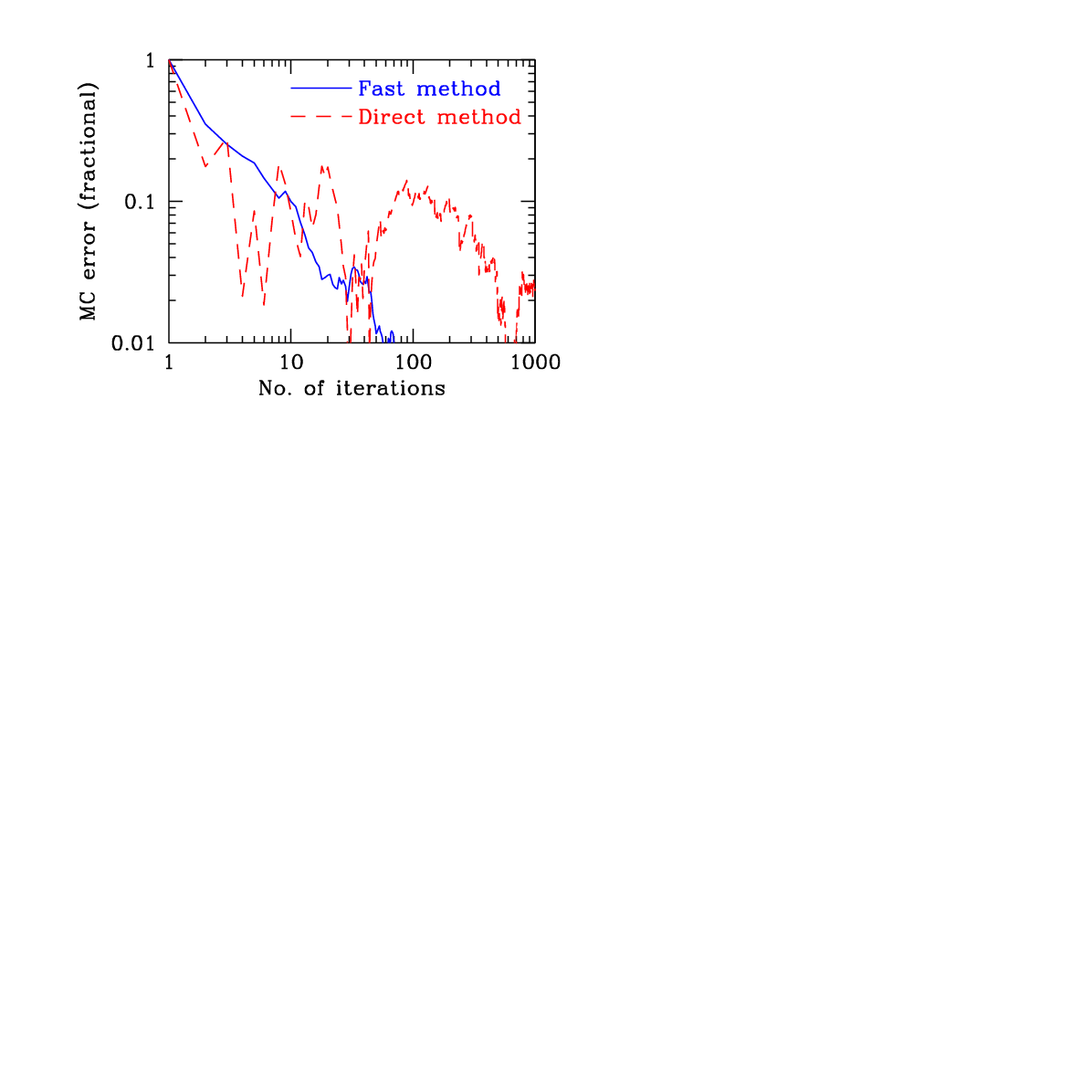}}
\caption{Monte Carlo error after $N$ simulations, compared for one instance of the ``fast'' Monte Carlo method
(Eq.~(\ref{eq:mcfancy})) and one instance of the ``direct'' method (Eq.~(\ref{eq:mcdirect})), showing much faster
convergence in the first case.
The local form of the bispectrum (Eq.~(\ref{eq:blocal})) was used, with
an all-sky homogeneous survey with $\ellmax=500$, noise level 1000 $\mu$K-arcmin, and beam $\theta_{\rm FWHM}=10$ arcmin.}
\label{fig:fastvsslow}
\end{figure}

We have found that the convergence of the first expression (Eq.~(\ref{eq:mcfancy})) is much more rapid
than the second (Eq.~(\ref{eq:mcdirect})).
This is illustrated in Fig.~\ref{fig:fastvsslow}, where we show the dependence of the error on the number of
Monte Carlo iterations, for both prescriptions.
It is seen that the Monte Carlo framework we have given above converges much more quickly than 
``direct'' Monte Carlo, requiring only $\sim 100$ simulations to reach 1\% accuracy.
For this reason, the Monte Carlo framework presented above is preferable to direct
Monte Carlo computation of the estimator variance, even though the computational cost per 
iteration is doubled (since $\nabla T$ must be computed instead of $T$, and two
$C^{-1}$ multiplications are needed instead of one).
Another benefit is that the linear term in the optimal estimator is computed in parallel.

It may seem suprising that the convergence rate of the two Monte Carlo prescriptions is so different.
One way to see this intuitively is to note that the first expression (Eq.~(\ref{eq:mcfancy})) contains two fewer powers
of the map $a$ than the second expression (Eq.~(\ref{eq:mcdirect})).
In effect, one factor of $(a a^T)$ has been replaced ``in advance'' by its Monte Carlo average $C$,
thus accelerating convergence.
(Note that the same phenomenon exists in the context of power spectrum estimation; if the Fisher matrix is computed by 
Monte Carlo trace evaluation as in Eq.~(\ref{eq:ohsmc}), the convergence rate will be much faster than estimating the covariance of
the estimator in Eq.~(\ref{eq:ps1}) by direct Monte Carlo.)


\section{Evaluation of $T$, $\nabla T$}
\label{sec:T}

We have now described a general Monte Carlo framework for optimal bispectrum estimation in
the presence of sky cuts and inhomogeneous noise, with two ingredients deferred: a method for
evaluating $\{T,\nabla T\}$, and a method for multipying a map by $(S+N)^{-1}$.
In this section, we address the first of these.  
We present a fast algorithm which, given an input set of multipoles $a=\{a_{\ell m}\}$ and bispectrum written in factorizable form
\be
b_{\ell_1\ell_2\ell_3} = \frac{1}{6} \sum_{i=1}^{\Nfact} X^{(i)}_{\ell_1} Y^{(i)}_{\ell_2} Z^{(i)}_{\ell_3} \symm
\label{eq:factyetagain}
\ee
evaluates $T[a]$ and $\nabla T[a]$.

If $X_\ell$ is any $\ell$-dependent quantity, define $X_a(\bx)$ to be the position-space map obtained
by applying the filter $X_\ell$ to $a_{\ell m}$:
\be
X_a(\bx) = \sum_{\ell m} X_\ell a_{\ell m} Y_{\ell m}(\bx)    \label{eq:Xadef}
\ee
The basis for our algorithm will be the following position-space expression for $T[a]$:
\be
T[a] = \frac{1}{6} \sum_{i=1}^{\Nfact} \int d\cos(\theta)\,d\varphi\,
                                  X^{(i)}_a(\theta,\varphi) Y^{(i)}_a(\theta,\varphi) Z^{(i)}_a(\theta,\varphi)
\label{eq:Tintegral}
\ee
obtained from Eq.~(\ref{eq:factyetagain}) and the identity
\bea
&& \int d\cos(\theta)\,d\varphi\, Y_{\ell_1 m_1}(\theta,\varphi) Y_{\ell_2 m_2}(\theta,\varphi) Y_{\ell_3 m_3}(\theta,\varphi) \nn \\
&& = \left[ \frac{(2\ell_1+1)(2\ell_2+1)(2\ell_3+1)}{4\pi} \right]^{1/2}   \nn \\
&& \qtwo \times \threej{\ell_1}{\ell_2}{\ell_3}{m_1}{m_2}{m_3} \threej{\ell_1}{\ell_2}{\ell_3}{0}{0}{0}. 
\eea
We next observe that the integral in Eq.~(\ref{eq:Tintegral}) can be done exactly using Gauss-Legendre quadrature in $\cos(\theta)$
with $\Ntheta = \lfloor 3\ellmax/2\rfloor + 1$ points, and uniform quadrature in $\varphi$ with $\Nphi = (3\ellmax+1)$ points.
(This observation is the basis for the GLESP pixelization \cite{Doroshkevich:2003xb}.)
This is because each term in the integrand is a polynomial in $\cos(\theta)$ of degree $\le 3\ell_{max}$, multiplied by a
factor $e^{im\varphi}$ with $-3\ellmax \le m \le 3\ellmax$.  (There is a subtlety here: some terms in Eq.~(\ref{eq:Tintegral}) 
contain odd powers of $\sqrt{1-z^2}$, for which Gauss-Legendre integration is not exact, but each such term has an odd
value of $m$, and hence gives zero when integrated over $\varphi$.)

This observation permits the integral to be replaced by a finite sum without approximation:
\be
T[a] = \frac{\pi}{3\Nphi} \sum_{i=1}^{\Nfact} \sum_{\theta,\varphi} w_\theta X^{(i)}_a(\theta,\varphi) 
                                                                             Y^{(i)}_a(\theta,\varphi) 
                                                                             Z^{(i)}_a(\theta,\varphi)
\label{eq:Tsum}
\ee
where $w_\theta$ denotes the Gauss-Legendre weight at the quadrature point $\theta$.

Our algorithm for evaluating $T[a]$, in the form~(\ref{eq:Tsum}), is given as follows.
There is an outer loop over $\theta$ in which the contribution of each isolatitude ring to the integral is accumulated.
Within this loop (i.e. for fixed $\theta$), the first step is to fill the matrix
\be
{\bf M}_{\ell m} = a_{\ell m} Y_{\ell m}(\theta,0)
\label{eq:step1}
\ee
using upward recursion in $\ell$ to generate the spherical harmonics.
Second, we evaluate the matrix
\be
{\bf N}_{\ell\varphi} = \sum_m {\bf M}_{\ell m} e^{i m \varphi}
\label{eq:step2}
\ee
by taking an FFT along each column of ${\bf M}$.
Third, we evaluate
\be
{\bf X}_{i\varphi} = \sum_\ell X^{(i)}_\ell {\bf N}_{\ell\varphi}
\label{eq:step3}
\ee
by matrix multiplication.
After this step, the matrix entry ${\bf X}_{i\varphi}$ is equal to the quantity $X^{(i)}_a(\theta,\varphi)$
defined in Eq.~(\ref{eq:Xadef}).
We compute matrices ${\bf Y}_{i\varphi}$, ${\bf Z}_{i\varphi}$ analagously,
replacing $X^{(i)}_\ell$ in Eq.~(\ref{eq:step3}) by $Y^{(i)}_\ell$, $Z^{(i)}_\ell$.
The fourth step is to accumulate the contribution of one isolatitude ring (in Eq.~(\ref{eq:Tsum})) to $T[a]$ as follows:
\be
T[a] \leftarrow T[a] + \frac{\pi w_\theta}{3\Nphi} \sum_{i,\varphi} {\bf X}_{i\varphi} {\bf Y}_{i\varphi} {\bf Z}_{i\varphi}\,.
\label{eq:step4}
\ee

This completes the algorithm for evaluating $T$; we now describe extra steps needed to evaluate $\nabla T$.
The idea is to compute derivatives of $T$ with respect to each of the quantities defined in Eqs.~(\ref{eq:step1})-(\ref{eq:step4}),
in reverse order.
First, differentiating Eq.~(\ref{eq:step4}), one computes
\be
\frac{\partial T}{\partial {\bf X}_{i \varphi}} = \frac{\pi w_\theta}{3\Nphi} {\bf Y}_{i\varphi} {\bf Z}_{i\varphi}  \label{eq:step5}
\ee
and likewise for $(\partial T/\partial {\bf Y}_{i\varphi})$ and $(\partial T/\partial {\bf Z}_{i\varphi})$.
Similarly, by differentiating Eqs.~(\ref{eq:step1})-(\ref{eq:step3}), one computes the following quantities in order:
\be
\frac{\partial T}{\partial {\bf N}_{\ell\varphi}} =
    \sum_i   X^{(i)}_\ell \frac{\partial T}{\partial {\bf X}_{i\varphi}}
                + Y^{(i)}_\ell \frac{\partial T}{\partial {\bf Y}_{i\varphi}}
                + Z^{(i)}_\ell \frac{\partial T}{\partial {\bf Z}_{i\varphi}}  \label{eq:step6}
\ee
\bea
\frac{\partial T}{\partial {\bf M}_{\ell m}}       &=&
    \sum_\varphi \frac{\partial T}{\partial {\bf N}_{\ell\varphi}} e^{im\varphi}  \label{eq:step7}  \\
\frac{\partial T}{\partial a_{\ell m}}             &=&
    \frac{\partial T}{\partial {\bf M}_{\ell m}} Y_{\ell m}(\theta,0)    \label{eq:step8}
\eea
The final result gives the contribution of one isolatitude ring to $\nabla_{\ell m}T = (\partial T/\partial a_{\ell m}^*)$.
This procedure computes $\{T[a],\nabla T[a]\}$ in twice the running time needed to compute $T[a]$ alone.

The algorithm we have just presented was closely inspired by the position-space estimator of 
Komatsu, Spergel and Wandelt \citep{Komatsu:2003iq}, in the context of the local bispectrum (Eq.~(\ref{eq:blocal})).
Indeed, our algorithm can be viewed as both generalizing the KSW estimator to an arbitrary input
bispectrum satisfying the factorizability condition, and permitting
calculation of the gradient $\nabla T[a]$ in addition to $T[a]$.
(We have seen that Monte Carlo evaluations of $\nabla T$ are needed to
compute the linear term in the estimator, the normalization $F_{\estE}$, and the variance $\Var(\estE)$
in the presence of sky cuts and inhomogeneous noise.)

Additionally, we have presented the details of the algorithm in an optimized way which dramatically improves
the running time, compared to existing implementations.
For example, in \cite{Creminelli:2005hu}, a running time of 60 CPU-minutes is quoted to evaluate the cubic term $T[a]$
for the local bispectrum (Eq.~(\ref{eq:blocal})), using a quadrature in the $r$ integral with 260 points,
at $\ell_{max}=335$.
With these parameters, our implementation evalues $T[a]$ in 27 CPU-{\em seconds}; after optimzing $\Nfact$
using the method of \S\ref{sec:optnfact}, this is further improved to 4 CPU-seconds.

The main reason that our implementation is so fast is that the only steps with cost $\bigoh(\Nfact\ellmax^3)$
have been written as matrix multiplications in Eqs.~(\ref{eq:step3}),~(\ref{eq:step6}).
These can be evaluated extremely efficiently using an optimzed library such as BLAS.
In existing implementations, the same asymptotic cost is accrued by means of $\Nfact$ separate $\bigoh(\ellmax^3)$
spherical harmonic transforms, but the overall prefactor is much larger.
A second, less important, optimization is that we have converted the integral (Eq.~(\ref{eq:Tintegral})) to a sum (Eq.~(\ref{eq:Tsum}))
using Gauss-Legendre quadrature in $\cos(\theta)$ and uniform quadrature in $\varphi$, rather than using a pixelization
such as Healpix.
In addition to giving an exact evaluation of Eq.~(\ref{eq:Tdef}), our ``pixelization'' is 
optimized to minimize the number of quadrature points needed, which translates 
to smaller matrix sizes in the rate-limiting steps.

\begin{table}
\begin{center}
\begin{tabular}{|c||c|c|}
\hline                                        &  WMAP3              &   Planck  \\
\hline                                        &  ($\ellmax=1000$)   &   ($\ellmax=2000$)  \\  \hline
Point source (Eq.~(\ref{eq:bps}))             &   3 CPU-sec         &   1 CPU-min  \\
ISW-lensing (Eq.~(\ref{eq:secbispec}))        &  10 CPU-sec         &   2 CPU-min  \\
Local (Eq.~(\ref{eq:Flocal}))                 &  84 CPU-sec         &  14 CPU-min  \\
Equilateral (Eq.~(\ref{eq:Feq})               & 117 CPU-sec         &  23 CPU-min  \\
Gravitational (Eq.~(\ref{eq:Fgrav}))          & 467 CPU-sec         &  81 CPU-min  \\
HD (Eq.~(\ref{eq:Fhd}))                       & 149 CPU-sec         &  38 CPU-min  \\ \hline
\end{tabular}
\end{center}
\caption{CPU time needed for one evaluation of $T[a]$, for each of the optimized bispectra from Tab.~\ref{tab:nopt}.
Evaluating $\nabla T[a]$ in addition to $T[a]$ would double the running times shown.}
\label{tab:ttimings}
\end{table}

In Tab.~\ref{tab:ttimings}, we have shown timings for one $T[a]$ evaluation in several mock surveys.
(Note that in the Monte Carlo framework for \S\ref{sec:est}, several such evaluations are needed in each
Monte Carlo iteration.)
Using the optimizations that we have presented here, which improve existing implementations by several orders of
magnitude, a fully optimal bispectrum analysis for Planck ($\ellmax=2000$) should easily be feasible.

\section{Simulating non-Gaussian maps}
\label{sec:ngsim}

So far, our emphasis has been on bispectrum estimation; however, another application of our machinery is that
it provides a fast algorithm for simulating non-Gaussian maps.
It should be emphasized from the outset that there is no ``universal'' probability distribution
for a field whose power spectrum and bispectrum are prescribed.
This is because the four- and higher-point connected correlation functions must be nonvanishing, for the
probability density to be positive definite.
In general, two schemes for simulating a non-Gaussian field, with the same power spectrum and bispectrum,
will differ in their higher-point amplitudes.
However, we expect that our algorithm will be useful in the regime of weak non-Gaussianity, where
higher-point amplitudes can be neglected.

We present a simulation algorithm which generates
all-sky simulated maps starting from arbitrary input power spectrum $C_\ell$, and
any input bispectrum $B_{\ell_1\ell_2\ell_3}$ which satisfies the factorizability condition (Eq.~(\ref{eq:factdef})).
The power spectrum and bispectrum of the field $a'_{\ell m}$ which is simulated will satisfy:
\bea
C'_\ell &=& C_\ell + \bigoh(B^2)                                   \label{eq:simCl} \\
B'_{\ell_1\ell_2\ell_3} &=& B_{\ell_1\ell_2\ell_3} + \bigoh(B^3)   \label{eq:simB}
\eea
where $\bigoh(B^k)$ denotes terms containing $k$ or more powers of the bispectrum.
For $N \ge 4$, the {\em connected} $N$-point function of the simulated field will satisfy:
\be
\langle a_{\ell_1m_1} a_{\ell_2m_2} \cdots a_{\ell_Nm_N} \rangle_{\rm conn.} = \bigoh(B^{N-2})    \label{eq:simhigher}
\ee
Under the assumption of weak non-Gaussianity, where the extra terms in Eq.~(\ref{eq:simB}) can be neglected, 
the power spectrum and bispectrum of the simulated field will agree with the input values $C_\ell$, $B_{\ell_1\ell_2\ell_3}$.
The problem of simulating non-Gaussian fields has received some attention in the literature
\citep{Komatsu:2003fd,Liguori:2003mb,Contaldi:2001wr,Rocha:2004ke}, but no method has been
proposed with this generality.

Our simulation algorithm is given as follows.  One first simulates a Gaussian random field $a_{\ell m}$ with power spectrum $C_\ell$.
Then define $a'_{\ell m}$ by perturbing to order $\bigoh(B)$ as follows:
\be
a'_{\ell m} = a_{\ell m} + \frac{1}{3} \nabla_{\ell m} T[ C_{\ell'}^{-1} a_{\ell'm'} ].
\label{eq:simdef}
\ee
The algorithm given in \S\ref{sec:T} is used to evaluate $\nabla T$.
The CPU time needed for one random realization of $a'_{\ell m}$ is therefore given by Tab.~\ref{tab:ttimings}
(with a factor of two included for calculating $\nabla T$ in addition to $T$), e.g. 168 CPU-seconds for the local
bispectrum at WMAP3 noise levels and $\ellmax=1000$.  
With this level of performance, non-Gaussian simulations can easily be included in a Monte Carlo analysis of Planck data,
in the generality of an arbitrary factorizable bispectrum.

To lowest (zeroth) order in $B$, the power spectrum of $a'_{\ell m}$ is $C_\ell$.
Let us calculate the lowest-order contribution to the bispectrum of $a'_{\ell m}$.
Plugging in the defintion (Eq.~(\ref{eq:gradT})) of $\nabla T$,
\bea
&& \!\!\! \langle a'_{\ell_1 m_1} a'_{\ell_2 m_2} a'_{\ell_3 m_3} \rangle  \nn \\
&& = \frac{1}{6} B_{\ell_1\ell'_2\ell'_3} \threej{\ell_1}{\ell'_2}{\ell'_3}{m_1}{m'_2}{m'_3} \times \\
&& \qquad \wick{12}{C_{\ell'_2}^{-1} <1 a^*_{\ell'_2 m'_2} C_{\ell'_3}^{-1} <2 a^*_{\ell'_3 m'_3} >1 a_{\ell_2 m_2} >2 a_{\ell_3 m_3}} \symm \nn \\
&& = B_{\ell_1\ell_2\ell_3}  \threej{\ell_1}{\ell_2}{\ell_3}{m_1}{m_2}{m_3}.
\eea
where $\symm$ denotes the sum over five additional terms obtained by permuting $(\ell_i,m_i)$.
This shows that the lowest-order bispectrum is simply $B_{\ell_1\ell_2\ell_3}$; it is easy to see that the 
orders of the higher-order terms are as claimed in Eq.~(\ref{eq:simB}).

%
\begin{figure}
\centerline{\epsfxsize=3.2truein\epsffile{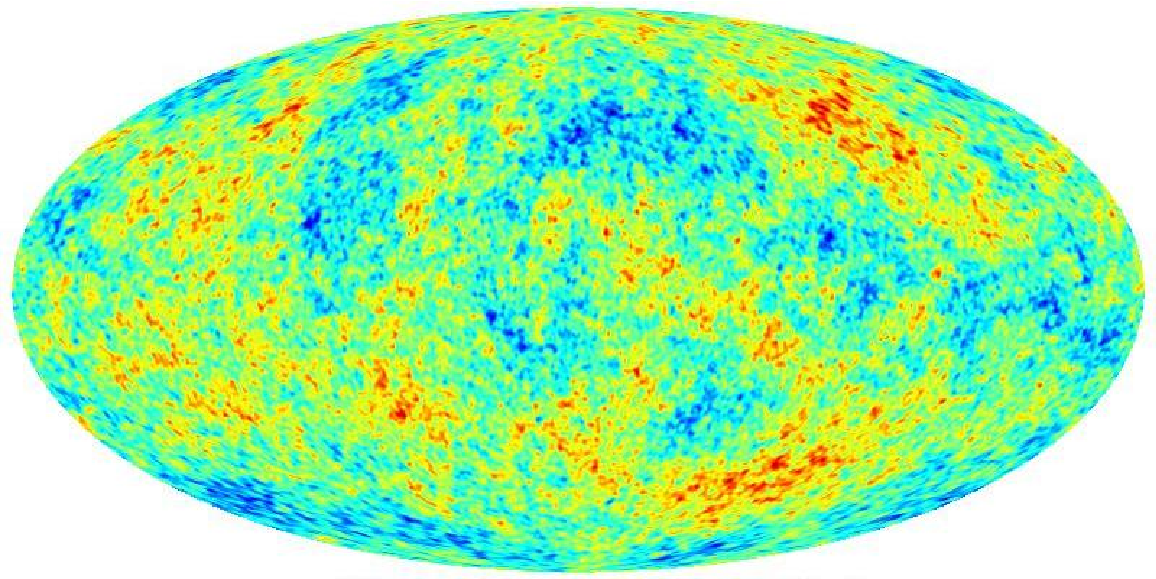}}
\centerline{\epsfxsize=3.2truein\epsffile{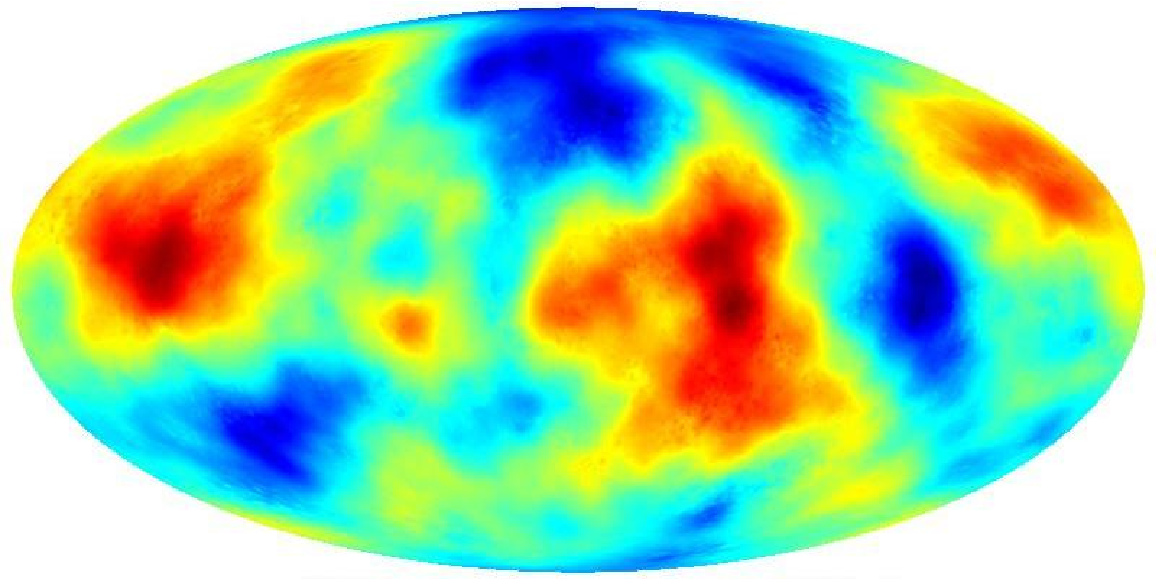}}
\centerline{\epsfxsize=3.2truein\epsffile{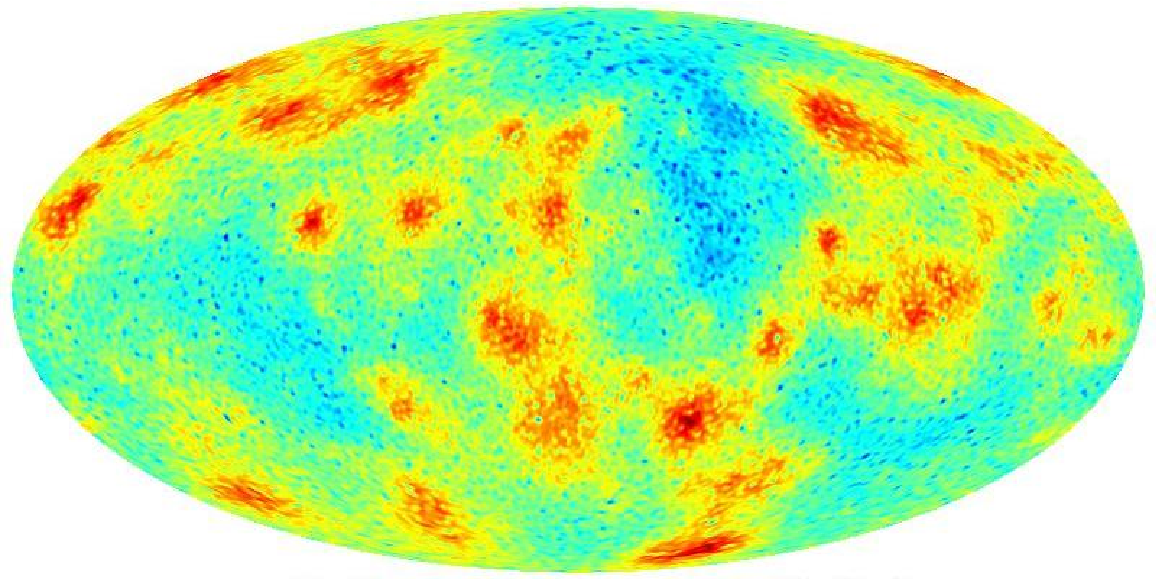}}
\caption{One random realization obtained using the simulation algorithm from \S\ref{sec:ngsim}, 
consisting of a Gaussian map $\{a_{\ell m}\}$ (top), and non-Gaussian maps $\{a^{\rm loc}_{\ell m}\}$ (middle), 
$\{a^{\rm eq}_{\ell m}\}$ (bottom).
To lowest order, the map $(a_{\ell m} + \fnlloc a^{\rm loc}_{\ell m} + \fnleq a^{\rm eq}_{\ell m})$
will have the power spectrum of the fiducial model, and bispectrum which is a linear combination of the local and 
equilateral forms (Eqs.~(\ref{eq:Flocal}),~(\ref{eq:Feq})).
All three maps have been smoothed with a $1^\circ$ Gaussian beam.}
\label{fig:syn}
\end{figure}

In Fig.~\ref{fig:syn}, we illustrate the method by showing a single random realization, split into three pieces.
From Eq.~(\ref{eq:simdef}), each realization is generated as a Gaussian piece (the first term on the right-hand side)
plus a small non-Gaussian perturbation (the second term) which depends on the bispectrum.
We have shown the Gaussian piece separately in Fig.~\ref{fig:syn}, and also shown the non-Gaussian term for the
case of the local (Eq.~(\ref{eq:Flocal})) and equilateral (Eq.~(\ref{eq:Feq})) bispectra.

Let us emphasize the caveats associated with the simulation algorithm presented here.
A model which predicts non-Gaussianity at the three-point level will also predict higher-point connected
correlation functions; these will not be reproduced faithfully by the simulations.
If higher-point correlations are important, then one must tailor the simulation method to the specific
model; one cannot expect to use a ``generic'' algorithm which only incorporates two-point and three-point
information, such as the algorithm we have given in this section.
However, in the regime of weak non-Gaussianity, where the three-point function is marginal and higher-point 
correlations are negligible, our simulation method should apply.

For the specific case of the local bispectrum $\Bloc_{\ell_1\ell_2\ell_3}$, the generic simulation algorithm has another caveat:
the non-Gaussian contribution to the power spectrum, while formally of order $\bigoh(f_{NL}^2)$, os unphysically
large even for moderate values of $f_{NL}$.
This is interpreted physically and discussed in more detail in the appendix of \cite{Hanson:2009kg}, where a
modification of the generic algorithm is proposed for the local shape, to eliminate the spuriously large non-Gaussian
power spectrum.
(For the case of the local shape, there is also an exact simulation algorithm which correctly simulates all
higher-point statistics, at the expense of somewhat increased computational cost \cite{Liguori:2003mb}.)

\section{Example surveys}
\label{sec:exsurveys}

The general Monte Carlo framework for optimal bispectrum estimation which
has been presented above applies to any survey for which a map can
be efficiently multiplied by $C^{-1}$.
We conclude by considering some mock surveys whose sky coverage and noise
are azimuthally symmetric.  The role of azimuthal symmetry is to make the
noise covariance matrix diagonal in $m$, so that $C^{-1}$ multiplication can
be performed quickly by ``brute force'' \cite{Oh:1998sr}.

\begin{figure}
\centerline{\epsfxsize=3.2truein\epsffile[50 500 320 700]{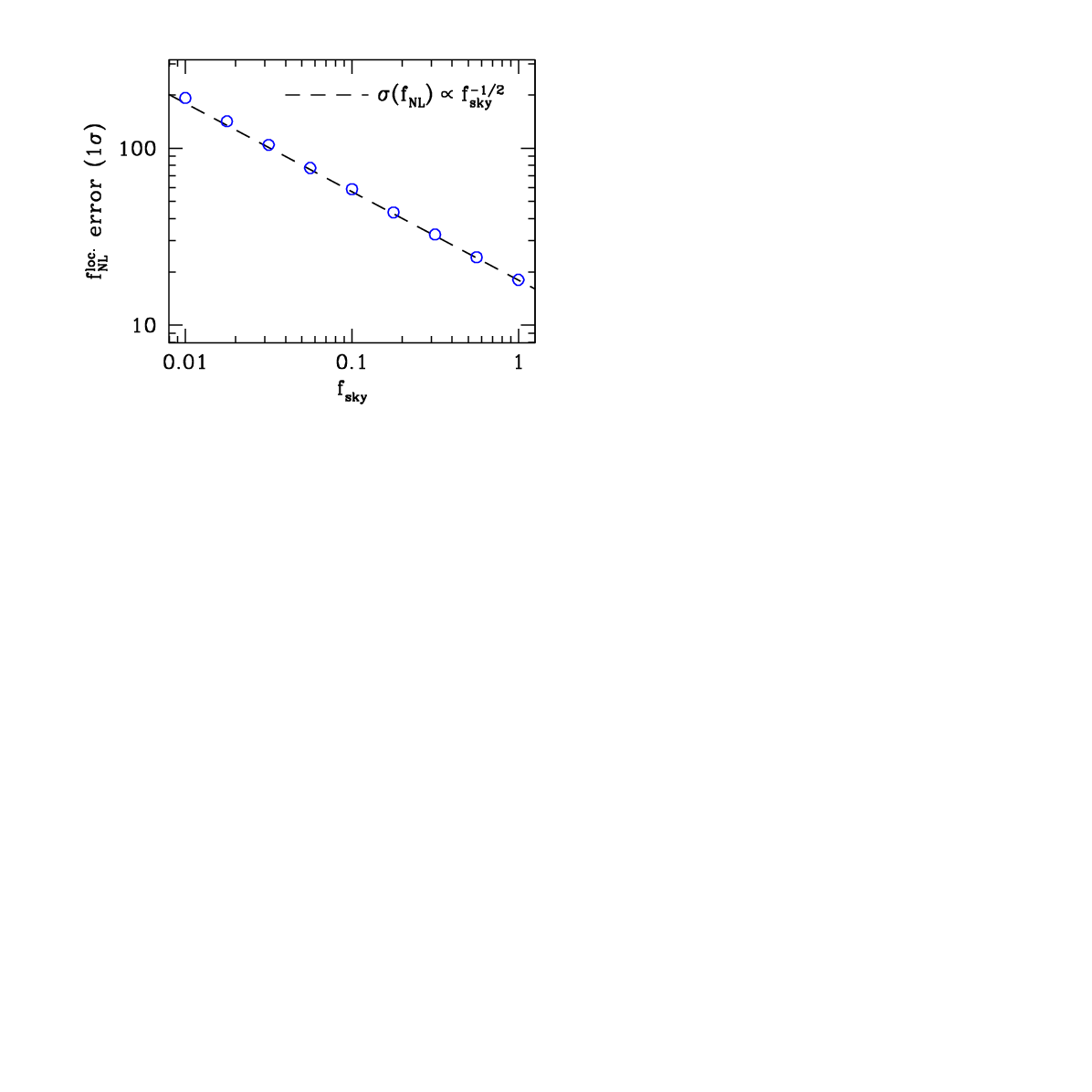}}
\caption{Dependence of the $1\sigma$ error $\sigma(\fnlloc)$ on $\fsky$, for a spherical cap shaped
survey with noise level 500 $\mu$K-arcmin and 10 arcmin beam, showing good agreement between the
Monte Carlo errors (circles), and simple $\fsky^{-1/2}$ scaling (dotted line).}
\label{fig:sigmavsfsky}
\end{figure}

Our first example will be a survey with homogeneous noise level 500 $\mu$K-arcmin
and Gaussian beam $\theta_{\rm FWHM}=10$ arcmin, with the geometry of a spherical cap.
In Fig.~\ref{fig:sigmavsfsky}, we show $1\sigma$ errors $\sigma(\fnlloc)$ obtained using the optimal estimation
framework from \S\ref{sec:est}, for varying $\fsky$.
It is seen that, with optimal estimators and for this simple sky cut, 
$\sigma(\fnlloc)$ varies as $\fsky^{-1/2}$ over two orders of magnitude.

\begin{figure}
\centerline{\epsfxsize=3.2truein\epsffile[50 500 320 700]{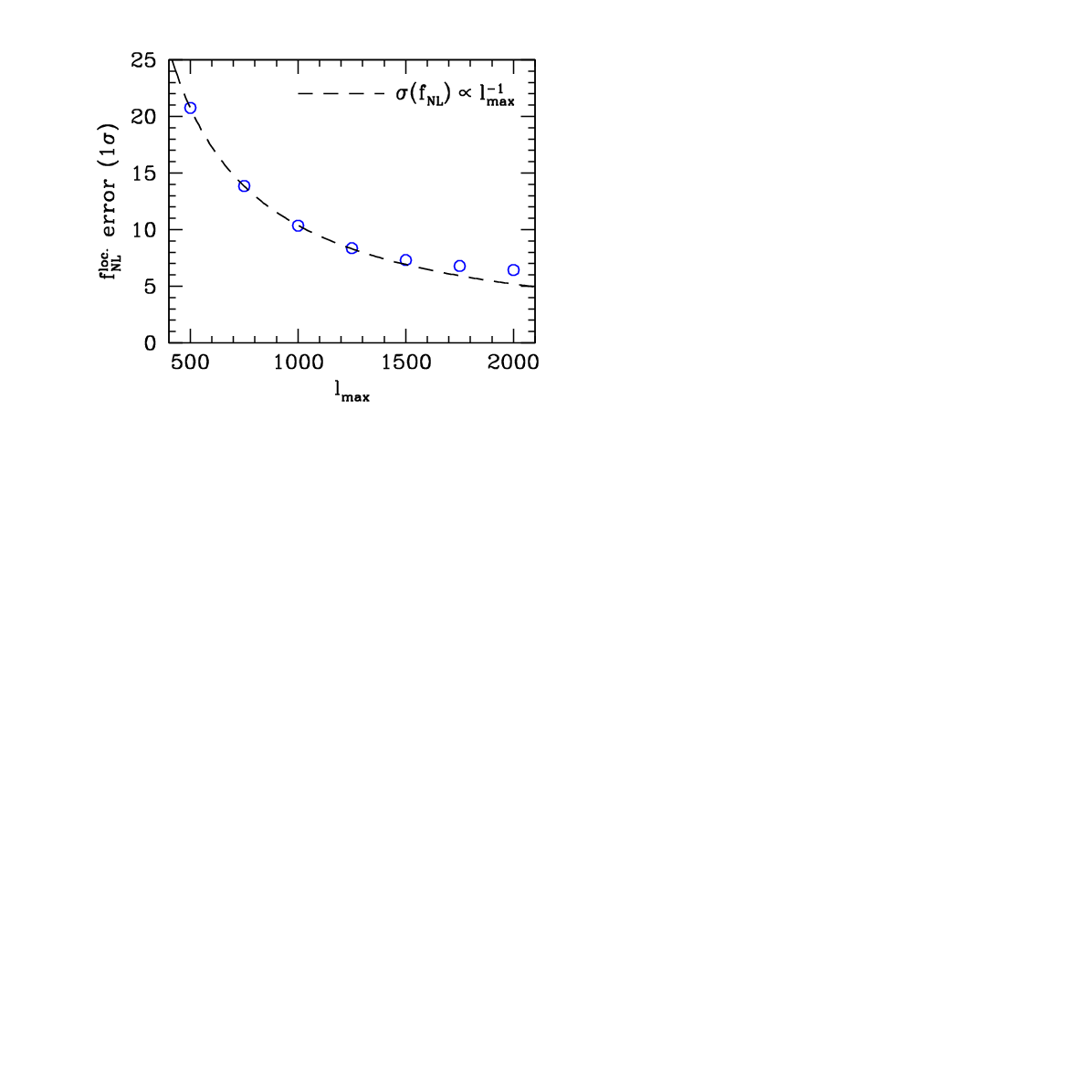}}
\caption{Dependence of the $1\sigma$ error $\sigma(\fnlloc)$ on $\ellmax$, for the azimuthally
symmetric approximation to Planck described in \S\ref{sec:exsurveys}, showing an $\ellmax^{-1}$
dependence throughout the sample variance limited regime.}
\label{fig:sigmavslmax}
\end{figure}

As a second example, we consider a Planck-like survey.
To approximate Planck within the constraint of azimuthal symmetry, we include a ``galactic'' sky cut 
which masks an equatorial band whose size is chosen to give $\fsky=0.8$, and take a pixel-dependent
noise variance of the form
\be
\sigma(\theta,\varphi) = \left( \frac{2}{\pi} \right) \sigma_0^2 \sin(\theta)
\ee
where the average noise level $\sigma_0$ is determined by the sensitivities in Tab.~\ref{tab:noise}.
(This angular dependence of the noise is motivated by the Planck scan strategy, which scans along
great circles through the ecliptic poles; however, note that azimuthal symmetry requires us to place
the poles perpendicular to the galactic cut.)

Since it is currently unclear what cutoff in $\ell$ will be needed in Planck to ensure that foreground 
contamination is sufficiently well-controlled for bispectrum estimation, we show the dependence of $\sigma(\fnlloc)$
on $\ellmax$ in Fig.~\ref{fig:sigmavslmax}.
It is seen that, throughout the range of multipoles where Planck is sample variance limited ($\ell \simle 1500$),
$\sigma(\fnlloc)$ varies roughly as $\ellmax^{-1}$, as expected from \cite{Komatsu:2001rj}.
At higher $\ell$, $\sigma(\fnlloc)$ flattens as instrumental noise becomes a contaminant.
At $\ellmax=2000$, the largest value considered, the $1\sigma$ error agrees with the Fisher matrix forecast from \S\ref{sec:fisherforecasts}.
This demonstrates the ability of the optimal estimator to saturate statistical limits on $\sigma(\fnlloc)$ in the presence
of sky cuts and inhomogeneous noise.
Another feature seen in Fig.~\ref{fig:sigmavslmax} is that $\sigma(\fnlloc)$ always decreases as $\ellmax$ increases.
This is a characteristic of optimal estimators, which can never worsen as more modes are added.
In contrast, for non-optimal bispectrum estimators, the variance eventually begins to worsen when $\ellmax$ becomes
large enough that not all modes are signal-dominated \cite{Spergel:2006hy,Creminelli:2006gc}.

\section{Discussion}
\label{sec:disc}

Perhaps the most fundamental problem in connecting a predicted shape of the CMB bispectrum with data
is that analysis techniques which allow an arbitrary angular bispectrum $B_{\ell_1\ell_2\ell_3}$ are
computationally prohibitive.
For example, even with the simplifying assumption of all-sky homogeneous noise, the cost of evaluating an
optimal estimator is $\bigoh(\ellmax^5)$, due to the number of nonzero terms in the harmonic-space sum (Eq.~(\ref{eq:Edef})).
In practice, this problem has meant that bispectrum estimation to date has been limited to a few special forms
of the bispectrum, such as the ``local'' shape (Eq.~(\ref{eq:Flocal})), where fast estimators are available.
On the other hand, there is a growing menagerie of theoretically motivated bispectra from secondary anisotropies 
\cite{Cooray:1999kg,Goldberg:1999xm,Hu:1999vq,Seljak:1998nu,Spergel:1999xn,Verde:2002mu} 
and early universe physics 
\cite{Alishahiha:2004eh,Chen:2006nt,Chen:2006xj,Rigopoulos:2005ae,Rigopoulos:2005us}, 
which one might wish to study in a dataset such as Planck.

We have shown that the factorizability criterion for bispectra (Eq.~(\ref{eq:factdef})) is a compromise in generality
which is specific enough to enable fast algorithms, yet general enough to encompass a wide
range (\S\ref{sec:fact}) of predicted shapes for $B_{\ell_1\ell_2\ell_3}$.
We have given several fast algorithms which operate in the generality of an arbitrary factorizable bispectrum.
In each case, the idea behind the algorithm is that the factorizability condition permits efficient
computation by translating from harmonic to position space.

The first such algorithm is an optimization algorithm which, given an input bispectrum written as a sum of
many factorizable terms, outputs an ``optimized'' bispectrum with fewer terms which closely approximates the input.
It is feasible to run this algorithm on very large input sizes; e.g. the higher-derivative bispectrum
for Planck (with $\Nfact = 80000$ and $\ellmax = 2000$) was treated, using a two-level optimization procedure
described in \S\ref{sec:optnfact}.
In this example, the optimization algorithm outputs a bispectrum which has only 86 factorizable terms, and is
provably indistinguishable from the input; the algorithm is only allowed to terminate when the Fisher distance 
between the two is $\le 10^{-6}$.

Second, we have given a general Monte Carlo based framework for optimal bispectrum estimation, which relies
on two ingredients: a method for computing the quantities $\{T[a], \nabla_{\ell m}T[a]\}$ defined in
Eq.~(\ref{eq:Tdef}), and a method for multiplying a map $a = \{a_{\ell m}\}$ by $(S+N)^{-1}$, where $S$ and $N$ are
signal and noise covariance matrices.
For the first of these, we have given a fast algorithm which can be thought of as 
generalizing the KSW estimator (constructed in \cite{Komatsu:2003iq} for the local shape) to any factorizable bispectrum,
and providing a convenient way to compute the linear term in the optimal estimator (Eq.~(\ref{eq:Edef})).
Additionally, we have described optimizations, such as rewriting the slowest steps as matrix multiplications,
which improve the running time of existing implementations by several orders of magnitude.
This speedup allows us to use large values of $\ellmax$ and should make a Monte Carlo analysis for Planck
very affordable (Tab.~\ref{tab:ttimings}).

Our estimator is fully optimal under the assumption that an affordable method can be found for multiplying a CMB
map by $(S+N)^{-1}$, where $S$ and $N$ are signal and noise covariances respectively.
Finding such a method will depend on the noise model and is outside the scope of this paper, where our emphasis
has on algorithmic aspects of bispectrum estimation and simulation which are not experiment-specific.
If no such method can be found, then our estimator is not fully optimal, but still includes the linear term,
which should improve constraints in the presence of inhomogeneous noise (App.~\ref{app:subopt}).
The ``$C^{-1}$ problem'' is a general ingredient in optimal estimators and also arises, e.g. for optimal
power spectrum estimation and for lens reconstruction.
We will discuss this problem, with emphasis on features of the noise model which arise in WMAP,
in a forthcoming paper analyzing three-year WMAP data.

Finally, we have given a simulation algorithm which generates random sky maps with prescribed power spectrum
and factorizable bispectrum.
This greatly extends the generality of existing methods for simulating non-Gaussian fields, and is computationally
inexpensive, easily permitting non-Gaussian simulations to be included in Monte Carlo based pipelines if needed.
An important caveat is that the higher-point correlation functions are not guaranteed to match the model which
gives rise to the bispectrum; the higher correlations are merely guaranteed to be small (Eq.~(\ref{eq:simhigher})).
Therefore, the simulation method should only be relied upon in the regime of weak non-Gaussianity, where higher
correlations are negligible.

As an application of these techniques, we have done a Fisher matrix forecast for multiple bispectra at Planck
sensitivities (\S\ref{sec:fisherforecasts}).
Of the bispectra considered, we found that four were nondegenerate: the point source, ISW-lensing, local, and equilateral shapes.
Correlations between these four are small, so that the shapes can be independently constrained.
However, at Planck sensitivity levels, the ISW-lensing bispectrum can still significantly bias estimates of
the local bispectrum, if not marginalized in the analysis.
We have also demonstrated the optimal estimator on example surveys (\S\ref{sec:exsurveys}), showing that it
is both computationally affordable and achieves statistical limits on $\sigma(\fnlloc)$
for a Planck-like survey with inhomogeneous noise and $\ellmax=2000$.

Our most general conclusion is that the factorizability criterion (Eq.~(\ref{eq:factdef})) is a promising approach for
bridging the gap between a theoretically motivated shape of the bispectrum and data.
We have described a generic ``toolkit'', with algorithms for Fisher forecasting, analysis, and simulation,
which can be implemented once and subsequently applied to any bispectrum which can be written in factorizable form.
Even if the number of factorizable terms appears to be intractably large ($\sim 10^5$), the optimization algorithm 
(\S\ref{sec:optnfact}) can still be used and may reduce the number of terms to a manageable level, as in the case of
the higher-derivative shape.

\section*{Acknowledgments}
We would like to thank Wayne Hu, Dragan Huterer, Michele Liguori, Eugene Lim and Bruce Winstein for useful discussions.
We acknowledge use of the FFTW, LAPACK, CAMB, and Healpix software packages.
KMS was supported by the Kavli Institure for Cosmological Physics through the grant NSF PHY-0114422.
MZ was supported by the Packard and Sloan foundations, NSF AST-0506556 and NASA NNG05GG84G.

\appendix

\section{Monte Carlo averages}
\label{app:mc}

The purpose of this appendix is to derive Eqs.~(\ref{eq:mcE}),~(\ref{eq:mcF}), and~(\ref{eq:estvar}) in \S\ref{sec:est},
in which Monte Carlo expressions are given for the linear term in the estimator $\estE$, the normalization $F_{\estE}$, 
and the variance $\Var(\estE)$.
We note that Monte Carlo averages involving $(C^{-1}a)$, where $a$ is a Gaussian random field with covariance $C$, can
be evaluated using the contraction:
\be
\wick{1}{ (<1C^{-1}a)_{\ell_1 m_1}  (>1C^{-1}a)_{\ell_2 m_2}} = C^{-1}_{\ell_1 m_1, \ell_2 m_2}.
\ee
It will be convenient to define the following quantities:
\bea
\alpha &\eqdef& B_{\ell_1\ell_2\ell_3} B_{\ell_4\ell_5\ell_6} 
                \threej{\ell_1}{\ell_2}{\ell_3}{m_1}{m_2}{m_3} \threej{\ell_4}{\ell_5}{\ell_6}{m_4}{m_5}{m_6}     \nn \\
       &&  \times   (C^{-1})_{\ell_1m_1,\ell_4m_4} (C^{-1})_{\ell_2m_2,\ell_5m_5} (C^{-1})_{\ell_3m_3,\ell_6m_6}  \nn \\
\beta_{\ell m} &\eqdef& B_{\ell\ell_2\ell_3} \threej{\ell}{\ell_2}{\ell_3}{m}{m_2}{m_3} C^{-1}_{\ell_2m_2,\ell_3m_3}
\eea
In terms of these, we evaluate the following Monte Carlo averages, using the definition (Eq.~(\ref{eq:Tdef})) of $T$:
\bea
\Big\langle \nabla_{\ell m} T[C^{-1} a] \Big\rangle_a 
            &=& \frac{1}{2} B_{\ell\ell_2\ell_3} \threej{\ell}{\ell_2}{\ell_3}{m}{m_2}{m_3}    \nn \\
            &&  \qquad \times \Big\langle (C^{-1}a)_{\ell_2 m_2} (C^{-1}a)_{\ell_3 m_3} \Big\rangle_a \nn  \\
            &=& \frac{1}{2} \beta_{\ell m}.
\label{eq:mc1}
\eea
\bea
&&  \Big\langle \nabla_{\ell_1 m_1} T[C^{-1}a] C^{-1}_{\ell_1 m_1, \ell_2 m_2}  \nabla_{\ell_2 m_2} T[C^{-1}a] \Big\rangle_a \nn  \\
&&   = \frac{1}{4} B_{\ell_1\ell_3\ell_4} B_{\ell_2\ell_5\ell_6} 
                    \threej{\ell_1}{\ell_3}{\ell_4}{m_1}{m_3}{m_4} \threej{\ell_2}{\ell_5}{\ell_6}{m_2}{m_5}{m_6} \nn \\
&& \qtwo \times C^{-1}_{\ell_1m_1,\ell_2m_2}  \Big\langle (C^{-1}a)_{\ell_3m_3} (C^{-1}a)_{\ell_4m_4}    \nn  \\
&& \qfour                                                 (C^{-1}a)_{\ell_5m_5} (C^{-1}a)_{\ell_6m_6} \Big\rangle_a  \nn \\
&&   = \frac{\alpha}{2} + \frac{1}{4} \beta_{\ell_1 m_1} C^{-1}_{\ell_1m_1,\ell_2m_2} \beta_{\ell_2m_2}.
\label{eq:mc2}
\eea
The Monte Carlo expression (Eq.~(\ref{eq:mcE})) for $\estE$ follows immediately from the 
definition (Eq.~(\ref{eq:Edef})) and Eq.~(\ref{eq:mc1}) above.

Turning next to the estimator normalization $F_{\estE}$, the definition (Eq.~(\ref{eq:Edef})) implies:
\bea
F_{\estE} &=& \frac{1}{6} B_{\ell_1\ell_2\ell_3} B_{\ell_4\ell_5\ell_6}
              \threej{\ell_1}{\ell_2}{\ell_3}{m_1}{m_2}{m_3} \threej{\ell_4}{\ell_5}{\ell_6}{m_4}{m_5}{m_6}  \nn \\
           &&  \times (C^{-1})_{\ell_1m_1,\ell_4m_4} (C^{-1})_{\ell_2m_2,\ell_5m_5} (C^{-1})_{\ell_3m_3,\ell_6m_6}  \nn \\
          &=& \frac{\alpha}{6}.
\eea
Comparing with Eq.~(\ref{eq:mc1}),~(\ref{eq:mc2}), the Monte Carlo expression (Eq.~(\ref{eq:mcF})) for $F_{\estE}$ follows.

Finally, from the definition (Eq.~(\ref{eq:Edef})), the estimator variance is given by a sum of three terms,
\bea
\Var(\estE[a]) &=& \frac{1}{F_\estE} \Big\langle T[C^{-1}a]T[C^{-1}a]   \label{eq:var1} \\
&& \qquad - T[C^{-1}a] (C^{-1}a)_{\ell m} \beta_{\ell m}                \nn \\
&& \qquad + \frac{1}{4} (C^{-1}a)_{\ell_1 m_1} \beta_{\ell_1 m_1} (C^{-1}a)_{\ell_2 m_2} \beta_{\ell_2 m_2} \Big\rangle_a  \nn
\eea
which are evaluated as follows:
\bea
&& \Big\langle T[C^{-1}a] T[C^{-1}a] \Big\rangle_a  \nn  \\
&&  = \frac{1}{36} B_{\ell_1\ell_2\ell_3} B_{\ell_4\ell_5\ell_6} 
        \threej{\ell_1}{\ell_2}{\ell_3}{m_1}{m_2}{m_3} \threej{\ell_4}{\ell_5}{\ell_6}{m_4}{m_5}{m_6}   \nn  \\
&&  \qquad \times \Big\langle (C^{-1}a)_{\ell_1 m_1} (C^{-1}a)_{\ell_2 m_2} (C^{-1}a)_{\ell_3 m_3}      \nn  \\
&&  \qthree                   (C^{-1}a)_{\ell_4 m_4} (C^{-1}a)_{\ell_5 m_5} (C^{-1}a)_{\ell_6 m_6} \Big\rangle_a  \nn  \\
&&  \qtwo = \frac{\alpha}{6} + \frac{1}{4} \beta_{\ell_1 m_1} C^{-1}_{\ell_1 m_1,\ell_2 m_2} \beta_{\ell_2 m_2}.
\eea
\bea
&& \Big\langle T[C^{-1}a] (C^{-1}a)_{\ell m} \beta_{\ell m} \Big\rangle_a  \nn \\
&& = \frac{1}{6} B_{\ell_1\ell_2\ell_3}  \threej{\ell_1}{\ell_2}{\ell_3}{m_1}{m_2}{m_3} \beta_{\ell m}  \nn \\
&& \qquad \times \Big\langle (C^{-1}a)_{\ell_1 m_1} (C^{-1}a)_{\ell_2 m_2} (C^{-1}a)_{\ell_3 m_3} (C^{-1}a)_{\ell m} \Big\rangle_a  \nn \\
&&  \qtwo  =  \frac{1}{2} \beta_{\ell_1 m_1} C^{-1}_{\ell_1 m_1,\ell_2 m_2} \beta_{\ell_2 m_2}.
\eea
\bea
&& \frac{1}{4} \Big\langle (C^{-1}a)_{\ell_1 m_1} \beta_{\ell_1 m_1} (C^{-1}a)_{\ell_2 m_2} \beta_{\ell_2 m_2} \Big\rangle_a  \nn  \\
&& \qtwo = \frac{1}{4} \beta_{\ell_1 m_1} C^{-1}_{\ell_1 m_1, \ell_2 m_2} \beta_{\ell_2 m_2}.  \label{eq:var2}
\eea
Putting Eqs.~(\ref{eq:var1})-(\ref{eq:var2}) together, we get
\be
\Var(\estE) = \frac{\alpha}{6F_{\estE}^2} = \frac{1}{F_{\estE}}
\ee
completing the derivation of Eq.~(\ref{eq:estvar}).

\section{No $C^{-1}$}
\label{app:subopt}

Our Monte Carlo framework for optimal bispectrum estimation (\S\ref{sec:est}) has one experiment-specific
requirement: a method for multiplying a CMB map by $C^{-1}$.
In this appendix, we consider the case where this operation is impractical.  We will see that
it is possible to preserve one feature of the optimal estimator:
improving the estimator variance by including the linear term in the bispectrum estimator.

We construct a bispectrum estimator by replacing $C^{-1}$ where it appears in the optimal estimator (Eq.~(\ref{eq:Edef}))
by some filter $F$ which approximates $C^{-1}$ as well as possible:
\be
\estE'[a] = \frac{1}{F_{\estE'}} \Big( T[Fa] - (Fa)_{\ell m} \Big\langle \nabla_{\ell m} T[Fa'] \Big\rangle_{a'} \Big)
\ee
where we use primes to distinguish this from the optimal estimator.
It can be shown that this choice of linear term minimizes the variance, if the cubic term $T[Fa]$ is assumed fixed.
(In particular, omitting the linear term from the estimator $\estE'$ defined above always worsens the variance.)
This estimator only depends on the ability to generate simulated signal + noise maps $(Fa)$ which are filtered
in the same way as the data.
Any realistic analysis pipeline can generate such Monte Carlo simulations.

By a calculation parallel to App.~\ref{app:mc}, it can be shown
the estimator normalization and variance are given by the following Monte Carlo averages:
\bea
F_{\estE'}  &=& \frac{1}{3} \Big\langle \nabla T[Fs] F \nabla T[C_\ell^{-1} s_{\ell m}] \Big\rangle_s  \label{eq:subopt1}  \\
\Var(\estE') &=& \frac{1}{3} \Big\langle \nabla T[Fa] FCF^T \nabla T[Fa] \Big\rangle   \label{eq:subopt2}       \\
            && \qquad - \frac{1}{3} \Big\langle \nabla T[Fa] \Big\rangle_a (FCF^T) \Big\langle \nabla T[Fa] \Big\rangle_a  \nn
\eea
Note that the two are not equal, in contrast to the optimal estimator.
In the first expression (Eq.~(\ref{eq:subopt1})), the Monte Carlo average is taken over signal-only realizations $s$
and the $C_\ell^{-1}$ simply refers to division by the signal power spectrum, without reference to the
noise model.
In the second expression (Eq.~(\ref{eq:subopt2})), the average is taken over filtered signal + noise realizations $(Fa)$ as usual.

The Monte Carlo prescription given in Eqs.~(\ref{eq:subopt1}),~(\ref{eq:subopt2}) 
does require multiplying CMB maps by $F$ and by $FCF^T$.
(Note that $FCF^T$ is just the covariance matrix of the simulated field $(Fa)$.)
These operations are easier than multiplying by $C^{-1}$, but in a ``pure'' Monte Carlo pipeline in which
the only possible operation is generating simulations, one can always fall back on direct Monte Carlo evaluations of
$\estE'$ to compute the normalization and variance.
(Note that computing the estimator normalization in this way requires non-Gaussian simulations, using the algorithm from \S\ref{sec:ngsim}.)
As discussed in \S\ref{ssec:directmc}, direct Monte Carlo will be slower than a scheme such as Eqs.~(\ref{eq:subopt1}),~(\ref{eq:subopt2}).

\vfill
\bibliography{newbispec}

\end{document}